\documentclass[aps,prb,showpacs,floatfix,twocolumn,showkeys,preprintnumbers,amsmath,amssymb]{revtex4}

\usepackage{graphicx,color}
\usepackage{dcolumn}

\usepackage[latin1]{inputenc}
\usepackage{color}
\usepackage{amsmath,amssymb}

\newcommand{\be}{\begin{equation}}
\newcommand{\ee}{\end{equation}}

\newcommand{\beq}{\begin{eqnarray}}
\newcommand{\eeq}{\end{eqnarray}}

\newcommand{\ba}{\begin{array}}
\newcommand{\ea}{\end{array}}

\newcommand{\ua}{\uparrow}
\newcommand{\da}{\downarrow}

\def\H1{\widehat{H}_1}

\renewcommand{\i}{\ensuremath{\mathrm{i}}}
\newcommand{\e}{\ensuremath{\mathrm{e}}}

\def  \be      {\hat{\bm e}}

\begin{document}


\title{Theory of spin-orbit coupling in bilayer graphene}

\author{S. Konschuh, M. Gmitra, D. Kochan and J. Fabian}

\affiliation{Institute for Theoretical Physics, University of Regensburg, 93040 Regensburg, Germany}
\date\today

\begin{abstract}

Theory of spin-orbit coupling in bilayer graphene is presented. The electronic
band structure of the AB bilayer in the presence of spin-orbit coupling and
a transverse electric field is calculated from first-principles using
the linearized augmented plane wave method implemented in the WIEN\emph{2k} code.
The first-principles results around the $\rm K$ points are fitted to a tight-binding model.
The main conclusion is that the spin-orbit effects in bilayer graphene derive essentially
from the single-layer spin-orbit coupling which comes almost solely from the $d$ orbitals.
The intrinsic spin-orbit splitting (anticrossing) around the $\rm K$ points is about $24\,{\rm \mu eV}$ for
the low-energy valence and conduction bands, which are closest to the Fermi level, similarly as in the single
layer graphene. An applied transverse electric field breaks space inversion
symmetry and leads to an extrinsic (also called Bychkov-Rashba) spin-orbit splitting.
This splitting is usually linearly proportional to the electric field. The peculiarity of graphene bilayer
is that the low-energy bands remain split by $24\,\rm{\mu eV}$ independently of the applied external field.
The electric field, instead, opens a semiconducting band gap separating these low-energy bands.
The remaining two high-energy bands are spin-split in proportion to the electric field; the proportionality coefficient
is given by the second intrinsic spin-orbit coupling, whose value is $20\,{\rm \mu eV}$. All the band-structure effects and their spin
splittings can be explained by our tight-binding model, in which the spin-orbit Hamiltonian is derived from symmetry considerations.
The magnitudes of intra- and interlayer couplings---their values are similar to the single-layer graphene ones---are determined
by fitting to first-principles results.
\end{abstract}

\maketitle

\section{Introduction}

Spin-orbit coupling is the most important interaction affecting electronic spin transport in nonmagnetic materials.
The use of graphene in spintronics \cite{Fabian2004:RMP,Fabian2007:APS} would require detailed knowledge of graphene's
spin-orbit coupling effects, as well as discovering ways of increasing and controlling them.
Bilayer graphene has attracted wide attention since it has a tunable gap caused by a transverse electric field
(gate).\cite{Min2007:PRB, Castro2007:PRL, Zhang2009:Nature} Although the gap seems to saturate at around 0.3\,eV, the possibility to turn on and off the
electronic transport makes bilayer graphene a suitable material for (mainly analog) electronic applications, including potentially spintronics ones.

The electronic band structure of bilayer graphene derives from that of single-layer graphene,
with taking into account for interlayer coupling. The basic electronic structure is well understood.\cite{Min2007:PRB, Nilson2008PRB} What is not known yet
is how, and by how much
the realistic spin-orbit interaction modifies the electronic
spectrum, especially when gated. The spin-orbit effects are important not only for
the fundamental electronic band structure and its topology, but also for understanding
such phenomena as spin relaxation (see the recent spin injection experiments in
Refs.~[\onlinecite{Yang2011:PRL, Han2011:PRL}]),
spin Hall effect,\cite{QiaoEtAl2011:PRL} magnetoanisotropy, or weak (anti)localization. Conventional
charge transport in bilayer graphene has been studied in detail,\cite{DasSarma2010:PRB}
but spin transport \cite{Ghosh2011:JAP} or spin-orbit induced charge
electronic transport \cite{Tse2011:PRB} in graphene bilayer is only starting to
be explored.

In this paper we argue that the spin-orbit coupling in bilayer graphene comes mainly
from the intra-layer spin-orbit coupling, contrasting earlier studies \cite{Guinea2010:arxiv, Hei-Wen-Lui2010:arxiv}
that predicted large interlayer effects, enhancing the spin-orbit spectral splittings
by a decade as compared with single layers. A recent model investigation \cite{Gelderen2010:PRB}
has reported on spectral features of a bilayer in the presence of both intra and interlayer
hopping spin-orbit parameters. The single-layer spin-orbit physics comes from
(nominally unoccupied) carbon atom $d_{xz}\pm \i d_{yz}$ orbitals \cite{Gmitra2009:PRB, Konschuh2010:PRB}
which hybridize with the $p_z$ ones. The $d$ orbitals spin-orbit coupling opens
a gap at the $\rm K$ points---the corners of the hexagonal Brillouin zone---of the value of about $24\,\rm{\mu eV}$.\cite{Gmitra2009:PRB, Abdelouahed2010:PRB}
The $\sigma$-$\pi$ hybridization, on which most studies
have focused, determines a further spin-orbit splitting in the presence of an external
transverse electric field,\cite{Min2006:PRB, Gmitra2009:PRB, Konschuh2010:PRB}
that can also arise from the substrate.\cite{Ertler2009:PRB}
The $\sigma$-$\pi$ hybridization comprises the Stark effect (shift of the $p_z$ orbitals in the presence
of an electric field), the on-site coupling of the shifted $p_z$ and $s$ orbitals,
and, finally, the spin-orbit splitting of the in-plane $p_x \pm \i p_y$ orbitals.

As is now common we call \emph{intrinsic} the spin-orbit splitting---which amounts
to spectral anticrossings at and around $\rm{K}$, while preserving the double
spin degeneracy---in the absence of a transverse
electric field, and \emph{extrinsic} (or Bychkov-Rashba
\cite{Bychkov1984:JETPL}) the spin splitting (lifting of the spin degeneracy) in the presence
of such a field.
The distinction is rather sharp in materials with a space inversion symmetry
(such as bilayer graphene), in which the electronic bands are always doubly degenerate;
application of an electric field breaks the space inversion symmetry and spin-orbit coupling removes
this degeneracy,\cite{Fabian2007:APS} usually in proportion to the field.
An additional source of extrinsic spin-orbit coupling are adatoms,\cite{CastroNeto2009:PRL}
which may cause spin relaxation in graphene by creating patches of enhanced spin-orbit
coupling.\cite{Ertler2009:PRB, Zhang2011:preprint, Dugaev2011:PRB}

This paper reports on comprehensive first-principles as well as tight-binding
investigations of the electronic structure of bilayer graphene, in the presence of spin-orbit coupling,
and in the presence (and absence) of an external transverse electric field. Furthermore, the most
generic spin-orbit Hamiltonian consistent with the $\rm{K}$ points symmetry is derived for bilayer graphene
in an external electric field. The first-principles
method we use is the linearized augmented plane wave technique with the generalized gradient approximation
\cite{Perdew1996:PRL} for the exchange-correlation potential, as embedded within the Wien$2k$
package.\cite{Blaha:Wien2k} The extended tight-binding model is constructed as an effective
single-orbital hopping model reduced from a multi-orbital tight-binding scheme, using only atomic
spin-orbit coupling. The tight-binding parameters are obtained by fitting
to the first-principles results around the $\rm{K}$ points. In the first-principles calculations
we take $1.42\,{\rm \AA}$ for the intra-layer atomic distance, and $3.35\,{\rm \AA}$ for the interlayer distance.
The vacuum layer is taken to be $20\,{\rm \AA}$, well enough to uncouple the bilayer in the
transverse direction in the three-dimensional periodic structure calculation.

Our main conclusion is that the single-layer spin-orbit coupling determines quantitatively
the spin-orbit induced anticrossings and spin splittings at and around $\rm{K}(\rm{K}')$. The
interlayer coupling of the two graphene sheets in the AB stacked bilayer produces
parabolic bands around the $\rm K$ points. Two bands remain close to the
Fermi level. These low-energy bands, one conduction and one valence, cross at $\rm K$. Spin-orbit
coupling leads to anticrossing of the two bands with the value of $24\,\rm{\mu eV}$, as in
single-layer graphene. This splitting is due to the presence of $d$ orbitals in the $\pi$-bands.
Removing $d$ (and higher) orbitals from our calculation, the gap is reduced to about
$1\,\rm{\mu eV}$, the typical value coming from the $\sigma$-$\pi$ hybridization. In a transverse
electric field an orbital gap opens, separating the conduction and valence bands. In addition,
spin-orbit coupling leads to spin splitting, removing the spin degeneracy at a given momentum.
This extrinsic splitting is peculiar in bilayers, due to interlayer orbital effects. At the
$\rm K$ points, the spin splitting is independent of the electric field (at typical field magnitudes),
with the value of $24\,\rm{\mu eV}$, given by the intrinsic splitting. Away from $\rm K$ the extrinsic
spin-orbit coupling begins to dominate, giving the splittings of roughly $10\,\rm{\mu eV}$ per field
of 1V/nm, increasing linearly with increasing field. The fine structure of the spin splittings
away from $\rm{K}$ is well described by the intra-layer spin-orbit couplings for the low-energy
conduction band. Quantitative fits to the low-energy valence band and high-energy bands
require introducing also
interlayer spin-orbit coupling parameters. This we do by deriving the most general
spin-orbit Hamiltonian at $\rm{K}$, which has 10 real parameters. By embedding this Hamiltonian
with the tight-binding scheme and fitting to our first-principles data we find that the
interlayer parameters are in magnitudes similar (about $10\,\rm{\mu eV}$) to the intra-layer
ones. Being off-diagonal, their actual contribution to the spectrum close to $\rm{K}$
is greatly suppressed.

This work is organized as follows. In section II we present the tight binding model including
the discussion of the relevant $d$ orbitals for a general $N$-layer AB-stacked graphene with intra-layer spin-orbit coupling.
This model is discussed in detail for bilayer graphene in section III. Next, in
section IV we present the first-principles results and the fits from the
tight-binding model around the $\rm K$ points. We discuss the intrinsic and extrinsic
spin-orbit splittings of the bands and the interlayer spin-orbit couplings.
Appendix A constructs the most generic effective spin-orbit Hamiltonian
for the graphene bilayer in an external electric field at the $\rm{K}(\rm{K}')$ point and, as well,
for an arbitrary momentum $\mathbf{k}$.

\section{Model Hamiltonian}

\subsection{Tight-binding Hamiltonian}

\begin{figure}[!h]
	\includegraphics[width=\columnwidth]{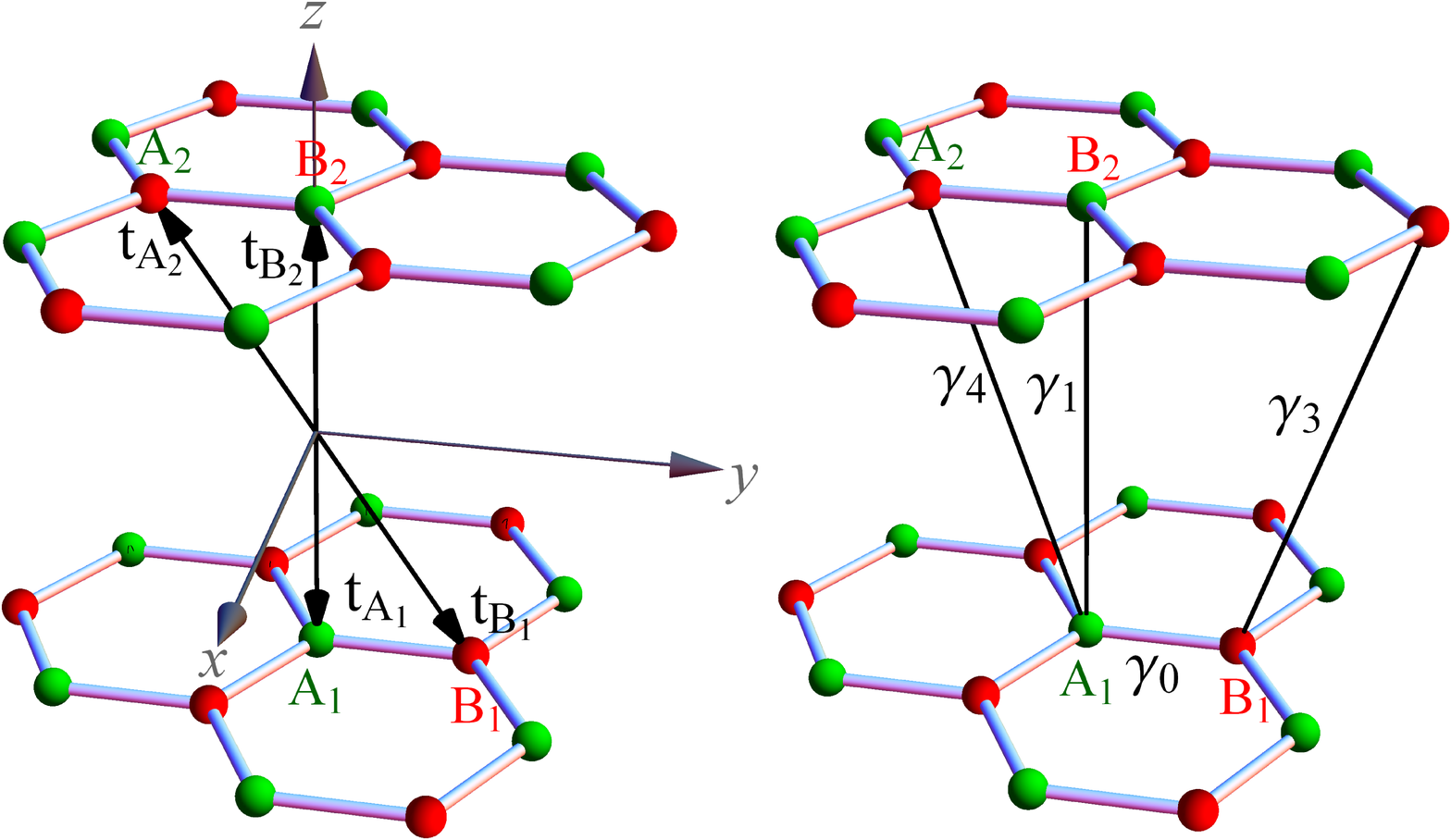}
	\caption{(Color online) Left: sketch of the AB stacked bilayer graphene.
                 The filled circles represent the carbon atoms on the sublattice (pseudospin) A and B, respectively.
                 Index 1 stands for the lower and index 2 for the upper carbon mono-layers.
                 Right: the hopping parameters $\gamma_0$, $\gamma_1$, $\gamma_3$, and $\gamma_4$
                 according to the Slonczewski-Weiss-McClure (SWMcC) convention. Atoms $\rm{B}_1$ and $\rm{B}_2$
                 are connected by $\gamma_4$ (not shown).
}
\label{BiStructure}
\end{figure}

The electronic structure of $\pi$-bands of graphite and of $N$-layers-graphene is usually described by a tight-binding approximation, often
parametrizied according to the Slonczewski-Weiss-McClure (SWMcC) model\cite{Slonczewski1958:PR,Johnson1972:PRB,Charlier1991:PRB,Charlier1992:PRB,Palser1999:PCCP,Latil2006:PRL,Partoens2006:PRB,Nilson2008PRB,Grueneis2008:PRB,Koshimo2009}
and expressed in terms of the $\pi$-band on-site orbital Bloch wave functions:
\begin{equation}\label{TB-basis}
\begin{aligned}
\Psi_{{\rm A}_i}(\mathbf{k})&=\frac{1}{\sqrt{N}}\sum\limits_{\mathbf{R}} \e^{\i\mathbf{k}(\mathbf{R}+\mathbf{t}_{{\rm{A}}_i})}\,p_{z}^{\rm eff}[\mathbf{r}-(\mathbf{R}+\mathbf{t}_{{\rm{A}}_i})]\,,\\
\Psi_{{\rm B}_i}(\mathbf{k})&=\frac{1}{\sqrt{N}}\sum\limits_{\mathbf{R}} \e^{\i\mathbf{k}(\mathbf{R}+\mathbf{t}_{{\rm{B}}_i})}\,p_{z}^{\rm eff}[\mathbf{r}-(\mathbf{R}+\mathbf{t}_{{\rm{B}}_i})]\,,
\end{aligned}
\end{equation}
labeled by quasi-momentum $\mathbf{k}$ counted from the $\rm{\Gamma}$ point, sublattice pseudospin $\rm A$ or $\rm B$, and the layer index
$i$, which runs form 1 to $N$ (the number of layers). Here, $\mathbf{t}_{{\rm{A}}_i}$ and $\mathbf{t}_{{\rm{B}}_i}$ stand for the positions of the
$2N$ atoms in the $N$-layer elementary cell (for the AB-stacked bilayer graphene the situation is depicted at Fig.~\ref{BiStructure}) and the summation
over $\mathbf{R}$ goes over all Bravais lattice vectors.

The intra- and interlayer hoppings between the (effective) $p_z$ orbitals of the neighboring atoms
are given by a set of parameters $\gamma$, schematically shown in Fig.~\ref{BiStructure}.
Parameters $\gamma_0$ and $\gamma_1$ describe the nearest neighbor intra-layer and interlayer hoppings, while $\gamma_3$ and $\gamma_4$ are indirect
hoppings between the layers.
In addition, the parameter $\Delta$ is introduced to handle the asymmetries in the energy shifts of the corresponding bonding and anti-bonding states
due to $\gamma_1$. The role of these hopping parameters in the band structure and the correspondence between the conventional tight-binding model and
the SWMcC parametrization is given in Ref.~[\onlinecite{Grueneis2008:PRB}].

Using the ordered on-site orbital Bloch basis $\Psi_{\rm{A}_1}(\mathbf{k})$, $\Psi_{\rm{B}_1}(\mathbf{k})$, $\Psi_{\rm{A}_2}(\mathbf{k})$, $\Psi_{\rm{B}_2}(\mathbf{k})$
the spinless $\pi$-band structure of the AB-stacked bilayer graphene with lower and upper layers placed in potential $V$ and $-V$, respectively, is described
by the effective $4\times 4\,$-\,Hamiltonian:
\beq\label{Heff_TB}
H_{\rm TB}(\mathbf{k})=\left(\ba{cccc}
	\Delta+V & \gamma_0 f(\mathbf{k}) & \gamma_4 f^*(\mathbf{k}) & \gamma_1 \\
	\gamma_0 f^*(\mathbf{k}) & +V & \gamma_3 f(\mathbf{k})& \gamma_4 f^*(\mathbf{k})\\
	\gamma_4 f(\mathbf{k})& \gamma_3 f^*(\mathbf{k}) & -V & \gamma_0 f(\mathbf{k})\\
	\gamma_1 & \gamma_4 f(\mathbf{k}) & \gamma_0 f^*(\mathbf{k})& \Delta -V
\ea\right).
\eeq
Here $f(\mathbf{k})$ is the nearest-neighbor structural function of the graphene hexagonal lattice with the lattice constant $a=2.46{\rm \AA}$:
\beq\label{TB-structural function}
f(\mathbf{k})=
\e^{\i\tfrac{a}{\sqrt{3}}k_y}\bigl[1 + 2\,\e^{-\i\tfrac{\sqrt{3} a}{2} k_y}\cos\bigl({\tfrac{a}{2}k_x}\bigr) \bigr]\,,
\eeq
which is accommodated to our on-site tight-binding basis (\ref{TB-basis}) and chosen coordinate system, see Fig.~\ref{BiStructure}.

With the spin degree of freedom $s=\{\ua,\da\}$ the on-site Bloch basis doubles:
\begin{equation}\label{TB-basis-spin}
\begin{aligned}
\Psi_{{\rm A}_i}(\mathbf{k}) &\mapsto \Psi_{{\rm A}_i,s}(\mathbf{k})=\Psi_{{\rm A}_i}(\mathbf{k})\otimes|s\rangle\,,\\
\Psi_{{\rm B}_i}(\mathbf{k}) &\mapsto \Psi_{{\rm B}_i,s}(\mathbf{k})=\Psi_{{\rm B}_i}(\mathbf{k})\otimes|s\rangle\,,\\
\end{aligned}
\end{equation}
and the dimension of the TB Hamiltonian increases to $4N\times 4N$:
\beq
H_{\rm TB}(\mathbf{k})\mapsto H_{\rm TB}(\mathbf{k})\otimes
\left(\begin{array}{cc}
\mid\ua\rangle\langle\ua\mid & 0\\
0 & \mid\da\rangle\langle\da\mid
\end{array}\right)\,.
\eeq
In what follows when using the on-site Bloch states for the $\rm{K}$ point momentum
$\mathbf{K}=\left(\tfrac{4}{3}\pi/a,0\right)$ we employ short-handed notation:
\beq\label{short-handed}
\Psi_{{\rm A}_i,s}=\Psi_{{\rm A}_i,s}(\mathbf{K})\,,\ \ \ \ \Psi_{{\rm B}_i,s}=\Psi_{{\rm B}_i,s}(\mathbf{K})\,.
\eeq

The $\pi$-bands on-site wave functions of few-layer graphene built solely on the $p_z$ orbitals are not affected by the atom's core spin-orbit
$\mathbf{L}\cdot\boldsymbol{s}$ term, since the $p_z$ orbitals carry zero orbital momentum. Therefore from the microscopical point of view, coupling of the
$p_z$ orbitals to other atomic orbitals is needed to describe spin-orbit effects. The minimum realistic model employs
$d_{\pm}=d_{xz} \pm \i d_{yz}$ orbitals, and also $s$ and $p_{\pm}=p_x \pm \i p_y$ orbitals if an external electric field is applied.
\cite{Gmitra2009:PRB} The resulting multi-orbital tight-binding model can be reduced by the L\"owdin transformation \cite{Lowdin1965:PR}
to obtain an effective Hamiltonian for the $\pi$-bands at the $\rm{K}$ point,\cite{Konschuh2010:PRB} where the normalized effective $p_z^{\rm{eff}}$ orbitals take the form:
\beq\label{pz-eff}
\begin{aligned}
p_z^{\rm{eff}}({\rm{A}}_i) &=\,\frac{1}{\sqrt{1+2\gamma^2}}\,\bigl[p_z({\rm{A}}_i)+\i\gamma\,
d_{+}({\rm{B}}_i)\bigr]\,,\\
p_z^{\rm{eff}}({\rm{B}}_i) &=\,\frac{1}{\sqrt{1+2\gamma^2}}\,\bigl[ p_z({\rm{B}}_i)+\i\gamma\,
d_{-}({\rm{A}}_i)\bigr]\,.
\end{aligned}
\eeq
The numerical value of the orbital mixing parameter $\gamma$ was estimated\cite{Konschuh2010:PRB} to be $0.09$.
For our group-theory based analyzes of the spin-orbit effects at the $\rm{K}(\rm{K}')$ point we do not need the explicit form of the effective
$p_z^{\rm eff}$ orbitals, the only information we need is that they transform as $\pi$-states. Exactly this requirement, as was already remarked
by Slonczewski,\cite{Slonczewski:thesis} implies that $p_z$ orbital centered at atom ${\rm{A}}_i$ (${\rm{B}}_i$) should be paired with $d_{+}$ ($d_{-}$)
orbital at atom ${\rm{B}}_i$ (${\rm{A}}_i$). However, as we will see later, the appearance of $d_{\pm}$
orbitals in $p_z^{\rm{eff}}$ is important for qualitative understanding of band spin splittings at $\rm{K}(\rm{K}')$.

\begin{figure*}
\includegraphics[width=1.8\columnwidth]{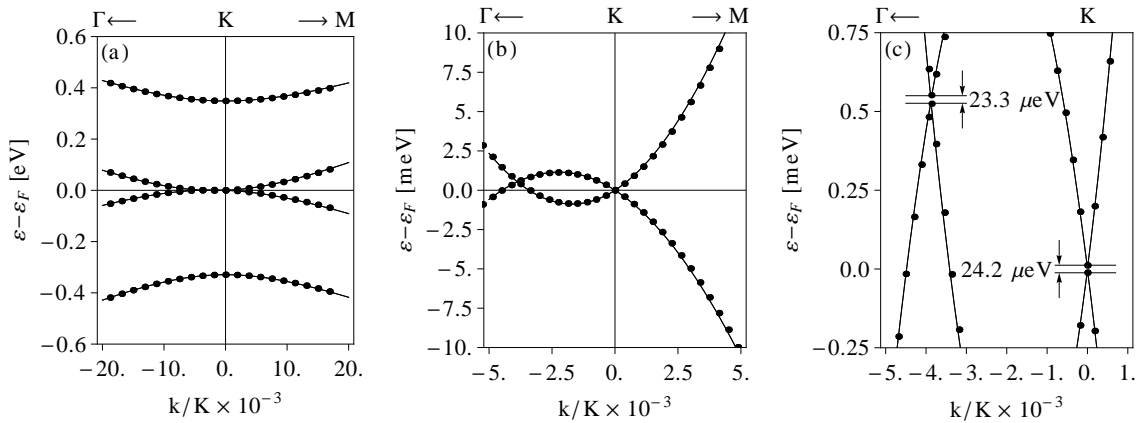}
\caption{Calculated band structure of the $\pi$-bands of bilayer graphene along
  the $\rm \Gamma KM$ lines, where $\mathrm{K=|\Gamma K|}=4\pi/(3a)$ with $a=2.46\,\rm \AA$.
  First-principles results are shown by circles while tight-binding calculations by solid lines.
  (a)~Low and high-energy bands.
  (b)~Fine structure of the low-energy bands.
  (c)~Detail view of the low-energy bands showing spin-orbit coupling induced anti-crossings
      at the $\rm K$ point and at $k=-0.063\,\rm nm^{-1}$.
 }
\label{bs0evpernm}
\end{figure*}

\subsection{Spin-orbit Hamiltonian}

The spin-orbit Hamiltonian at the $\rm{K}(\rm{K}')=\left(\pm \tfrac{4}{3}\pi/a,0\right)$ point in the presence of an external transverse electric field
is derived in detail from the group-theory arguments in Appendix A.
It possesses single-layer-like \textit{intrinsic} and \textit{extrinsic} (Bychkov-Rashba) spin-orbit couplings, whose strengths are given by the
four intrinsic $\lambda_{\rm I1}$, $\lambda_{\rm I1}'$, $\lambda_{\rm I2}$, $\lambda_{\rm I2}'$ and two extrinsic
$\overline{\lambda}_0\pm 2\lambda_{\rm BR}$ parameters, respectively. Their physical meanings and importance will be explained later.
Additionally, the symmetry group at $\rm{K}$($\rm{K}'$) allows four interlayer spin-orbit parameters $\lambda_1$, $\lambda_3$ and
$\overline{\lambda}_4\pm\delta\lambda_4$, whose indices refer to the spin-orbit interlayer geometry analogous to the SWMcC hopping convention.
Within the basis of the on-site spin Bloch functions (\ref{TB-basis-spin}) the spin-orbit at $\rm{K}(\rm{K}')$ Hamiltonian is given as follows:
\begin{widetext}
\beq\label{Heff_SOC}
H_{\rm SO}=\left(\ba{cccc}
	\tau\lambda_{\rm I2} s_z & \i(\overline{\lambda}_0 + 2\lambda_{\rm BR})s_{-}^{\tau} & \i(\overline{\lambda}_4+\delta\lambda_4)s_{+}^\tau & \tau\lambda_1 s_z \\
	-\i(\overline{\lambda}_0 + 2\lambda_{\rm BR})s_{+}^{\tau} & -\tau\lambda_{\rm I1} s_z & \i\lambda_3 s_{-}^\tau & -\i(\overline{\lambda}_4-\delta\lambda_4)s_{+}^\tau\\
	-\i(\overline{\lambda}_4+\delta\lambda_4)s_{-}^\tau & -\i\lambda_3 s_{+}^\tau & \tau\lambda_{\rm I1}' s_z & -\i(\overline{\lambda}_0 - 2\lambda_{\rm BR}) s_{-}^{\tau} \\
	\tau\lambda_1 s_z & \i(\overline{\lambda}_4-\delta\lambda_4)s_{-}^\tau & \i(\overline{\lambda}_0 - 2\lambda_{\rm BR})s_{+}^{\tau} & -\tau\lambda_{\rm I2}' s_z
   \ea\right)\,.
\eeq
\end{widetext}
In the above expression $\tau=1(-1)$ for the $\rm{K}(\rm{K}')$ point, respectively, and the matrices $s_z$ and $s_{\pm}^{\tau}=\tfrac{1}{2}(s_x\pm \i\tau s_y)$
stand for the $z$-component and raising and lowering spin operators. With zero electric field the bilayer graphene has inversion symmetry and
the number of spin-orbit $\lambda$-parameters reduces from ten to four, see the discussion in Appendix A, or Ref.~[\onlinecite{Guinea2010:arxiv}];
these are $\lambda_{\rm{I}1}=\lambda_{\rm{I}1}'$, $\lambda_{\rm{I}2}=\lambda_{\rm{I}2}'$, $\overline{\lambda}_{0}$ and $\overline{\lambda}_{4}$ and all others
are forced to be zero from symmetry requirements.
The spin-orbit couplings and hence $H_{\rm{SO}}$ are momentum
independent. However, at the end of App.~A we discuss their possible $\mathbf{k}$-dependent extension, which, as we will see, plays a very minor role for the
spectra near the $\rm{K}(\rm{K}')$ point.
Therefore the full model Hamiltonian for gated bilayer graphene in the vicinity of the $\rm{K}(\rm{K}')$ point is:
\beq\label{Heff_bi}
H_{\rm eff}(\mathbf{k})=H_{\rm TB}(\mathbf{k})\otimes\left(\begin{array}{cc}
\mid\ua\rangle\langle\ua\mid & 0\\
0 & \mid\da\rangle\langle\da\mid
\end{array}\right)
+H_{\rm SO}\,.
\eeq
The resulting electronic spectra of bilayer graphene derived from Hamiltonian (\ref{Heff_bi}) in the presence of a
transverse electric field, as well as our first-principles results, are presented in the following sections.

\section{Bilayer graphene}

The electronic bands of bilayer graphene around the $\rm K$ point are parabolic,
in contrast to (in the absence of spin-orbit coupling) linear bands in single-layer
graphene. When gated, a tunable band gap opens.\cite{OOstinga2008:NatureMaterials,Zhang2009:Nature,Min2007:PRB}
It has recently been proposed that the effects of spin-orbit coupling in bilayer
graphene are of the order of hundreds of $\rm{\mu eV}$, caused by effective
spin-dependent interlayer hopping between $p$ orbitals.\cite{Guinea2010:arxiv,Hei-Wen-Lui2010:arxiv}
Our results presented below do not support this view.

\subsection{Summary of results}


Figure~\ref{bs0evpernm} shows the calculated electronic band structure of bilayer graphene
around the $\rm K$ point along the $\rm{\Gamma KM}$ high-symmetry lines; spin-orbit coupling is taken
into account, but there is no applied electric field so the spin degeneracy is present.
There is excellent agreement between the tight-binding model and the first-principles calculations,
in all the energy scales shown. Each scale in the figure has its own physics.
Figure~\ref{bs0evpernm}(a) shows the usual picture of four, spin-degenerate, parabolic $\pi$-bands.
The high-energy conduction and valence bands are formed mainly from the $p_z$ orbitals at atoms
${\rm A_1}$ and ${\rm B_2}$ at the $\rm K$ point (see Fig.~\ref{BiStructure}). These
bands are shifted in energy by about
340$\,{\rm meV}$ by the direct interlayer hopping $\gamma_1$ away from the low-energy
bands, formed predominantly by the orbitals at atoms ${\rm A_2}$ and ${\rm B_1}$. The two
low-energy bands, again one conduction and one valence, are closest to the Fermi level.
In the tight-binding model the difference in the energy shift between the conduction
and valence bands is taken into account by the parameter $\Delta$. Although we do
not explicitly specify so, the spectra at the $\rm{K}'$ points are identical and our discussion
is valid also for them.

\begin{figure}
\includegraphics[width=\columnwidth]{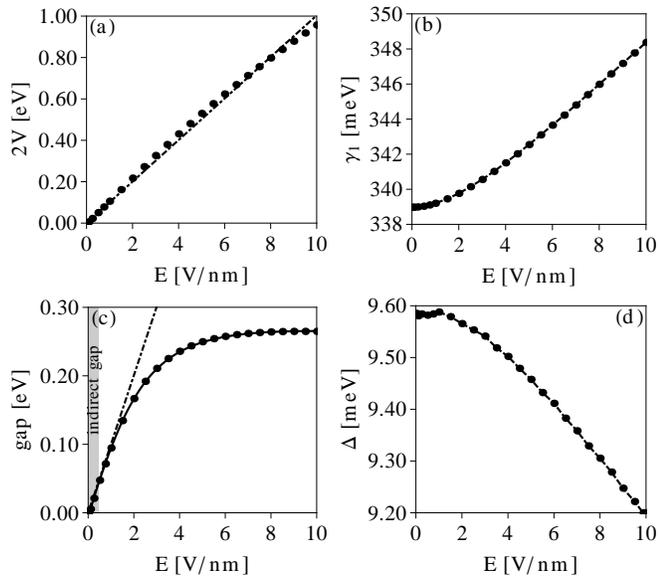}
\caption{Bilayer graphene essentials in an external electric field obtained from
first-principles calculations (circles). The figure
plots the electric field dependencies of
(a)~the electrostatic potential $2V$; the slope is described by the effective
interlayer distance of $d_{\rm eff}=0.1~\rm nm$,
which is defined by $2V=eEd_{\rm eff}$; the dashed line here is the linear fit.
(b)~the hopping parameter $\gamma_1$ obtained by fitting to the first-principles data at $\rm K$;
(c)~the energy gap in biased bilayer graphene, compared to the voltage $2V$ (dashed-dotted line);
the solid line here is the tight-binding calculation using the potential $2V$ from
the first-principles data in (a);
(d)~the parameter $\Delta$ obtained by fitting to the first-principles data at $\rm K$. Note that $E$ is the actual external electric field and
not the screened one, as presented for example in Ref.~[\onlinecite{Min2007:PRB}].}
\label{Efielddependence}
\end{figure}

\begin{figure}
\includegraphics[width=0.7\columnwidth]{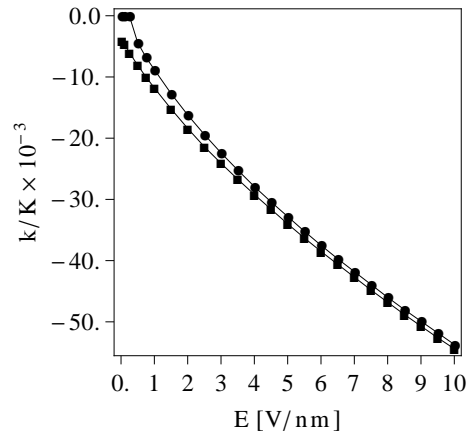}
\caption{The positions of the lattice momenta along the $\rm{\Gamma K}$ line (measured from $\rm K$),
corresponding to the minimum of the conduction band (circles) and to the maximum of the valence band (squares),
for different values of the transverse electric field. The first-principles and tight-binding results coincides.}
\label{maxandmin}
\end{figure}

Figure~\ref{bs0evpernm}(b) reveals a fine structure of the low-energy bands. The bands
form two overlapping parabolas, crossing at the $\rm K$ point, directly at the Fermi level,
as well as at the point of accidental crossing along the $\rm{\Gamma K}$ line, at about
0.5 meV above the Fermi level. This is the manifestation of the trigonal warping, that induces a
breaking of the Fermi surface in the vicinity of each Dirac point into four pockets (Lifshitz transition),
see for example [\onlinecite{Lemonik2010:PRB}].These crossings are governed
by the indirect interlayer hopping parameters $\gamma_3$ and $\gamma_4$, which pull the two bands
together.
The spin-orbit coupling causes anti-crossings of $24\,{\rm \mu eV}$, just as
in single-layer graphene,\cite{Gmitra2009:PRB} as seen in Fig.~\ref{bs0evpernm}(c).
The anti-crossings collapse below $1\,{\rm \mu eV}$ if $d$ and higher orbitals are excluded from the calculations.

\begin{figure*}
\includegraphics[width=1.8\columnwidth]{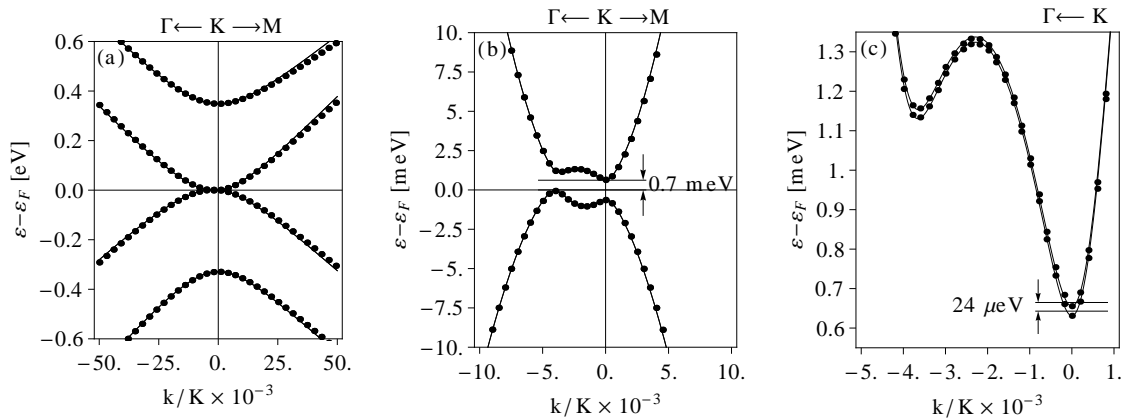}
\caption{Calculated band structure of the $\pi$-bands of bilayer graphene
in a transverse electric field of $25\,\rm{mV/nm}$. The spectra along
the $\rm{\Gamma KM}$ line are shown, with the
$\rm K$ point and the Fermi energy at the origin; $\mathrm{K=|\Gamma K|}=4\pi/(3a)$ with $a=2.46\,\rm \AA$.
The first-principles calculations are shown by circles, while the lines come from tight-binding modeling.
(a)~Low and high-energy bands.
(b)~The fine structure of the low-energy bands with the corresponding indirect band gap of
   $0.7\,\rm meV$ between $k=-0.068\ \rm nm^{-1}$ and the $\rm K$ point.
(c)~Detail view at the low-energy conduction band split by extrinsic spin-orbit coupling;
   the maximum value of the splitting of $2\lambda_{\rm I}= 24\,{\rm\mu eV}$ is at the $\rm K$ point and at
   $k=-0.063\ \rm nm^{-1}$.
 }
\label{bs0.025evpernm}
\end{figure*}

Applying a transverse external electric field $E$ to a bilayer places the two layers
at a different electrostatic potential. In the tight-binding model this is described by
introducing a potential $2V$, which includes all possible screening effects and
corresponds to the splitting of the low-energy bands at the K point.\cite{Min2007:PRB}
Figure \ref{Efielddependence}(a) shows $2V$, extracted from the first-principles calculations,
as a function of the electric field. The dependence is almost linear with the slope
of about 0.1 electron nanometers, corresponding to the effective electrostatic bilayer
distance $d_{\rm eff} \approx  0.1$ nm. The electric field induces also a slight variation of
the parameters $\Delta$ and of the direct interlayer hopping $\gamma_1$, obtained by
fitting to the tight-binding model. The corresponding dependencies are shown
in Figs~\ref{Efielddependence}(b) and \ref{Efielddependence}(d). In Fig.~\ref{Efielddependence}(c)
we plot the spectral gap as a function of the electric field. At low electric fields the
band gap is manifestly indirect, with a marked difference between the minimum of the conduction
band and the maximum of the valence band. As the electric field increases beyond 1 V/nm, the
gap becomes almost direct. The lattice momenta of the conduction band minimum and the valence
band maximum are plotted in Fig.~\ref{maxandmin} for reference.
The corresponding physics of the gap opening is discussed below.

\begin{figure*}
\includegraphics[width=1.8\columnwidth]{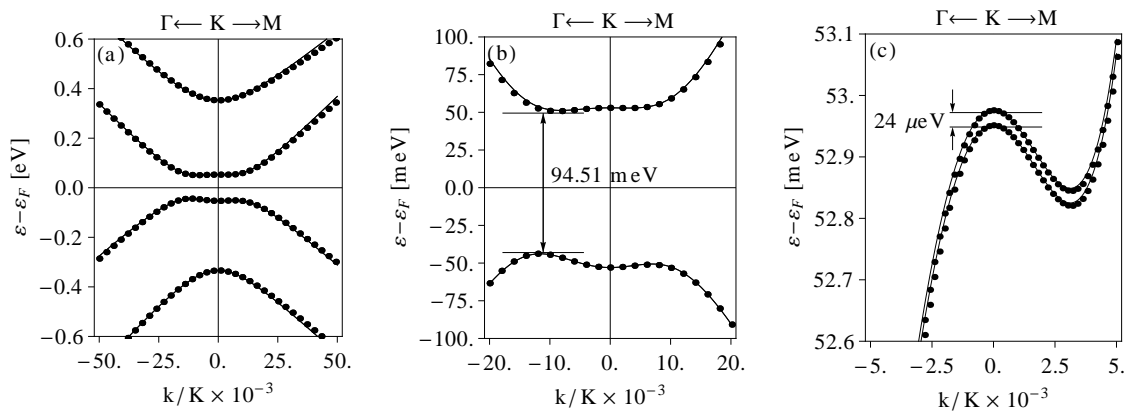}
\caption{Calculated band structure of the $\pi$-bands of bilayer graphene with the applied
electric field of $1\,\rm{V/nm}$. Circles show the results of the first-principles and
lines of the tight-binding calculations.
(a)~Low and high-energy bands, displaying a bandgap.
(b)~View of the low-energy bands showing the (mostly) direct bandgap of $94.5\,$ meV
   between the valence (at $k=-0.2\ \rm nm^{-1}$) and the conduction (at $k=-0.15\ \rm nm^{-1}$) bands.
(c)~Detail view of the low-energy conduction band showing the spin splitting,
    with the maximum value of $2\lambda_{\rm I}= 24\,{\rm\mu eV}$ at the $\rm{K}$ point and in its close vicinity.
 }
\label{bs1evpernm}
\end{figure*}

At small electric fields, less then $6\,$mV/nm, bilayer graphene is a
semimetal. A finite Fermi surface of a triangular shape
is formed from the low-energy bands; the electric field induces only
small energy gaps at the crossing points of the two overlapping parabolas (manifestation of trigonal warping).
A further increase of the electric field
opens an indirect band gap between the maximum of the valence band present at
the $\rm{K\Gamma}$ line and the minimum of the conduction band present at the $\rm{K}$ point (Lifshitz transition).
The corresponding band structure is shown in Fig.~\ref{bs0.025evpernm}. The global
picture of the bands is seen in Fig.~\ref{bs0.025evpernm}(a), while the opening
of the indirect band gap of 0.7 meV is seen in the fine structure zoom in Fig.~\ref{bs0.025evpernm}(b).

\begin{figure*}
\includegraphics[width=1.8\columnwidth]{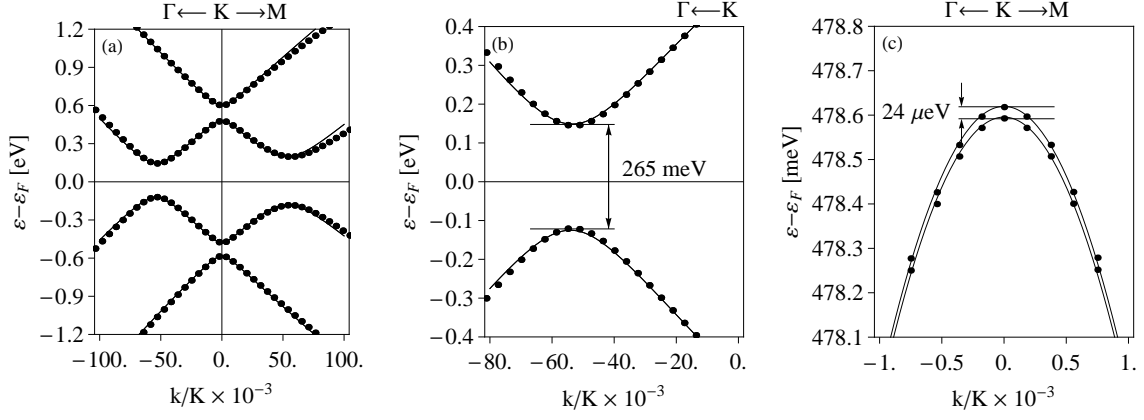}
\caption{Calculated band structure of the $\pi$-bands of bilayer graphene with the applied
electric field of $10\,\rm{V/nm}$. Circles show the results of the first-principles and
lines of the tight-binding calculations.
(a)~Hyperbolic high-energy bands and the Maxican-hat structure of the low-energy bands.
(b)~The fine structure of the low-energy bands in the bandgap region with the (mostly) direct
   bandgap of $265\,\rm meV$; the maximum of the valence band is at $k=-0.910\ \rm nm^{-1}$ and the
   minimum of the conduction band is at $k=-0.904\ \rm nm^{-1}$.
(c)~Detail view at the low-energy conduction band shows its splitting due spin-orbit coupling with
   the value of $2\lambda_{\rm I}=24\,{\rm\mu eV}$ at the $\rm{K}$ point.
 }
\label{bs10evpernm}
\end{figure*}

At electric fields greater than $0.45\,\rm{V/nm}$ the states at the $\rm{K}$ point repel significantly;
the minima and the maxima of the valence and conduction bands are present along the
$\rm{\Gamma K}$ and $\rm{KM}$ lines, and not at the $\rm{K}$ points.\cite{Gelderen2010:PRB,Min2007:PRB}
The band gap becomes (mostly) direct; the minimum of the conduction low-energy band and the maximum
of the valence low-energy band are in general at very close but still different momenta.
The spectrum for the electric field of 1 V/nm is shown in
Fig.~\ref{bs1evpernm}(a), with the fine structure showing the opening of the
direct gap of 94.5 meV in Fig.~\ref{bs1evpernm}(b). Finally,
an extreme case, that of a very high-field of 10 V/nm, is shown
in Fig.~\ref{bs10evpernm}. The direct band gap of 265 meV is seen in Fig.~\ref{bs10evpernm}(b).
Due to the shift of the conduction band minimum away from the $\rm{K}$ point, the bandgap at \
fields greater than $0.45\,$V/nm
is no longer proportional to the potential $2V$ (which determines the splitting at the
$\rm{K}$ point), but rather it saturates to a value
of about $265\,{\rm meV}$ \cite{Ohta2006:Science,Castro2007:PRL,Zhang2009:Nature}
as is shown in Fig.~\ref{Efielddependence}(c).

The applied electric field breaks space inversion symmetry and lifts the spin degeneracy.
The spin splittings for the low-energy conduction bands round the $\rm K$ points for $E=25\,$mV/nm,
$E=1\,$V/nm, and $E=10\,$V/nm are shown in Figs~\ref{bs0.025evpernm}(c), \ref{bs1evpernm}(c),
and \ref{bs10evpernm}(c), respectively. At these large field ranges, the spin splitting
at the $\rm{K}$ point is seen to be independent of the field, having a constant value of the
intrinsic splitting of $24\,{\rm \mu eV}$, deriving from the $d$ orbitals.

As seen from Figs~\ref{bs0evpernm}, \ref{bs0.025evpernm}, \ref{bs1evpernm}, and
\ref{bs10evpernm}, all the spectra around the $\rm K$ points, including the fine structures and the
spin splittings, can be faithfully described by tight-binding modeling. In the
following we analyze the spin-orbit-coupling induced anticrossings and spin splittings.

\subsection{Analysis of results}

\subsubsection{ Spin-orbit coupling at and near the $\rm{K}$ point}

The spin-orbit effects at $\rm K$ can be understood essentially in terms of the interplay between
the electrostatic potential $2V$, direct interlayer hopping $\gamma_1$, and \textit{intrinsic}
spin-orbit couplings controlled by $\lambda_{\rm I1}$, $\lambda_{\rm I1}'$, $\lambda_{\rm I2}$, $\lambda_{\rm I2}'$.
Basically with only those parameters, the energy spectrum of bilayer graphene at the $\rm{K}$ point ordered from high to low, reads
\begin{eqnarray}
\label{spectrumatKone}
\varepsilon_1^{\ua} & = & \Delta + \sqrt{\gamma_1^2 + V^2}+ \frac{\gamma_1\lambda_1+(\lambda_{\rm{I2}}+\lambda_{\rm{I2}}')\,V/2}{\sqrt{\smash[b]{\gamma_1^2  + V^2}}}, \\
\varepsilon_1^{\da} & = & \Delta + \sqrt{\gamma_1^2 + V^2}- \frac{\gamma_1\lambda_1+(\lambda_{\rm{I2}}+\lambda_{\rm{I2}}')\,V/2}{\sqrt{\smash[b]{\gamma_1^2  + V^2}}}, \\
\varepsilon_2^{\da} & = & + V + \lambda_{\rm I1}, \\
\varepsilon_2^{\ua} & = & + V - \lambda_{\rm I1}, \\
\varepsilon_3^{\ua} & = & - V + \lambda_{\rm I1}', \\
\varepsilon_3^{\da} & = & - V - \lambda_{\rm I1}', \\
\varepsilon_4^{\da} & = & \Delta - \sqrt{\gamma_1^2 + V^2}+ \frac{\gamma_1\lambda_1+(\lambda_{\rm{I2}}+\lambda_{\rm{I2}}')\,V/2}{\sqrt{\smash[b]{\gamma_1^2  + V^2}}}, \\
\label{spectrumatKeight}
\varepsilon_4^{\ua} & = & \Delta - \sqrt{\gamma_1^2 + V^2}- \frac{\gamma_1\lambda_1+(\lambda_{\rm{I2}}+\lambda_{\rm{I2}}')\,V/2}{\sqrt{\smash[b]{\gamma_1^2  + V^2}}},
\end{eqnarray}
as sketched in Fig.~\ref{energyspectrum}.
The above spectrum can be derived from the Hamiltonian (\ref{Heff_bi}) when treating the spin-orbit interaction
in the first order perturbation theory.

The values for $2V$, $\gamma_1$, $\Delta$, \emph{intrinsic} spin-orbit couplings $\lambda_{\rm I1}$, $\lambda_{\rm I1}'$, $\lambda_{\rm I2}$,
$\lambda_{\rm I2}'$ and direct interlayer spin-orbit parameter $\lambda_1$ are obtained by comparing the eigenvalues
Eq.(\ref{spectrumatKone}-\ref{spectrumatKeight}) of the effective tight-binding bilayer Hamiltonian, Eq.(\ref{Heff_bi}), to the first-principles spectra
at the $\rm{K}$ point. This analysis shows that the spin splittings of the low-energy
valence and conductance bands at the $\rm{K}$ point are the same and do not depend on the applied electric field; the spin-orbit parameters for these bands are
predicted to be $2\lambda_{\rm I1}\simeq 2\lambda_{\rm I1}'=24\,\rm{\mu eV}$.
In contrast, the spin splittings of the high-energy valence and conductance bands at the $\rm{K}$ point depend on the applied electric field. However, the
high-energy \emph{intrinsic} spin-orbit couplings governing these splittings are field independent and their values are fixed by
$2\lambda_{\rm I2}\simeq 2\lambda_{\rm I2}'=20\,\rm{\mu eV}$. Finally the direct spin-dependent interlayer parameter $\lambda_1=0$.
The remaining hopping ($\gamma_3$ and $\gamma_4$) and spin-orbit ($\overline{\lambda}_0$, $2\lambda_{\rm BR}$, $\lambda_3$, $\overline{\lambda}_4$ and $\delta\lambda_4$) parameters, as discussed in the next subsection, are chosen to reproduce the band-structure in the vicinity of the $\rm{K}$ point
(see the largest ranges shown in Figs. \ref{bs0evpernm}, \ref{bs0.025evpernm}, \ref{bs1evpernm}, and \ref{bs10evpernm}).
The parameters for selected values of electric field used in this paper are listed in Tab.~\ref{parameterstable}.

\begin{table*}
\begin{tabular}{l|cccccc|cccccccc|r}\hline\hline
                  &                             & {}                         & {}          & {}                         &                          & {}                      & {}                          & {}                      &                 &                &                           &                          &           & \\
TB                & $\Delta$	                & $\gamma_0$  & $\gamma_1$	 & $\gamma_3$  & $\gamma_4$	   & $2V$ \  \  & \ \ $2\lambda_{\rm I1}$\ & \ $2\lambda_{\rm I2}$\  & \ $\overline{\lambda}_0$\   & \ $2\lambda_{\rm BR}$\  & \ $\lambda_1$\  & \ $\lambda_3$\ & \ $\overline{\lambda}_4$\ & \ $\delta\lambda_4$\ & \,$\rm{SO}$  \\
SWMcC             & $\Delta-\gamma_2+\gamma_5$	& $\gamma_0$  & $\gamma_1$	 & $\gamma_3$  & $-\gamma_4/2$ & $2V$ \ \   &                          & {}                      & {}                          & {}                      & {}              & {}             & {}                        & {}                       &              \\
                  &                             & {}                         & {}          & {}                         &                          & {}                      & {}                          & {}                      &                 &                &                           &                          &          & \\
\hline
                  &                             & {}                         & {}          & {}                         &                          & {}                      & {}                          & {}                      &                 &                &                           &                          &   & \\
$E=0$			  & 0.0097	                    & 2.6 	      & 0.339		 & 0.28		   & -0.140	       & 0    \ \   & \ \ 24                   & 20                      & 5                           & 0                       & 0               & 0              & -12                       & 0                        &   \\
$E=25\,\rm mV/nm$ & 0.0096	                    & 2.6  	      & 0.339		 & 0.28		   & -0.145	       & 0.0013 \ \ & \ \ 24                   & 20                      & 5                           & 0.25                    & 0               & 0.038          & -12                       & -0.075                   &   \\
$E=1\,\rm V/nm$	  & 0.0096	                    & 2.6  	      & 0.339		 & 0.25		   & -0.165 	   & 0.1059 \ \ & \ \ 24                   & 20                      & 5                           & 10                      & 0               & 1.5            & -12                       & -3                       &   \\
$E=6\,\rm V/nm$	  & 0.0094	                    & 2.6  	      & 0.343		 & 0.29		   & -0.143 	   & 0.6238 \ \ & \ \ 24                   & 20                      & 5                           & 60                      & 0               & 9              & -12                       & -18                      &   \\
$E=10\ \rm V/nm$  & 0.0092	                    & 2.6 	      & 0.348		 & 0.26		   & -0.100 	   & 0.9572 \ \ & \ \ 24                   & 20                      & 5                           & 100                     & 0               & 15             & -12                       & -30                      &   \\
                  &                             & {}                         & {}          & {}            &        \ \ &                          & {}                      & {}                          & {}                      &                 &                &                           &                          &   \\
\hline\hline
\end{tabular}
\caption{Tight-binding (TB) parameters in the units of $\rm eV$s and spin-orbit (SO) couplings in the units of $\rm \mu eV$s,
obtained by fitting the band structure to the first-principles calculations.
The signs of the TB parameters are chosen to be consistent with the Slonczewski-Weiss-McClure (SWMcC)
parametrization,\cite{Dresselhaus1980:AP} which is also shown. The translation table of the
parameters in the tight-binding and SWMcC models is obtained from band-structure fitting
of graphite. The presented values of the TB parameters are similar to those
found elsewhere,\cite{Partoens2006:PRB,Grueneis2008:PRB,Zhang2008:PRB,FanZhang2010:PRB}
and are consistent with the values of Ref.~[\onlinecite{Min2007:PRB}] obtained
from bilayer band-structure calculation using the WIEN$2k$ code.
}
\label{parameterstable}
\end{table*}

Within the first order perturbation theory at the $\rm{K}$ point the eigenstates of the Hamiltonian (\ref{Heff_bi}) can be expressed
(apart from the overall normalization) in terms of the on-site spin Bloch wave-functions (\ref{TB-basis-spin}), for the form of the
eigenstates see Fig.~\ref{energyspectrum}, where
\beq\label{alpha}
\alpha_{\pm}=\frac{\sqrt{\smash[b]{\gamma_1^2+V^2}}\pm V}{\gamma_1}\,.
\eeq

{\begin{figure}
\includegraphics[width=\columnwidth]{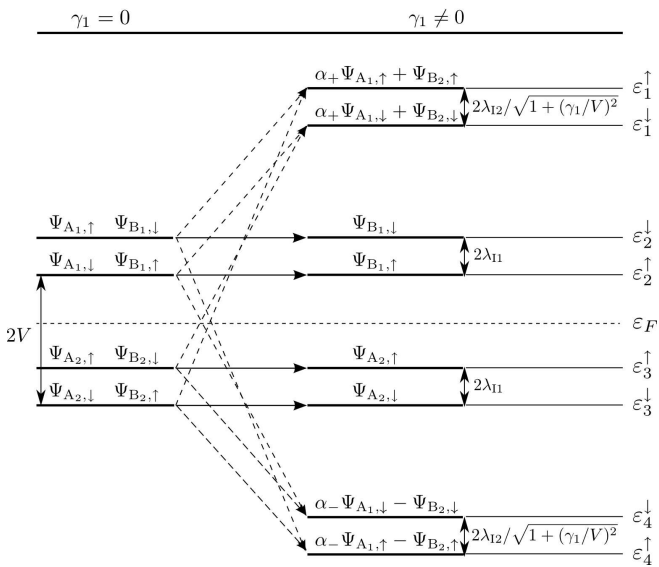}
\caption{Energy spectrum diagram of the AB stacked bilayer graphene at the $\rm{K}$ point.
The diagram at left shows the spin-orbit energetics of two interlayer-non-interacting ($\gamma_1=0$)
graphene sheets placed in the transverse potential difference $2V$.
The diagram at right represents the spin-orbit energetics in the presence of a transverse external field
including the direct interlayer interaction (in TB Hamiltonian mediated via the parameter $\gamma_1$).
The states formed predominantly by atoms $\rm A_2$ and $\rm B_1$ are split by the intrinsic spin-orbit coupling
$2\lambda_{\rm{I}1}$. These states form the low-energy valence and conductance bands. The states residing mainly on
$\rm A_1$ and $\rm B_2$ form the high-energy bands and are shifted in the energy spectrum by the direct interlayer
hopping $\gamma_1$. They are spin split by $2\lambda_{\rm I2}/\sqrt{\smash[b]{1+(\gamma_1/V)^2}}$.
The energies $\varepsilon_i^{s}$ ($i=\{1,2,3,4\}$ and $s=\{\ua,\da\}$) and the corresponding eigenstates are ordered from top
to bottom; see Eqs.\,(\ref{spectrumatKone}-\ref{spectrumatKeight}).
}
\label{energyspectrum}
\end{figure}}

To understand qualitatively the bilayer spectrum and the spin splittings at the $\rm{K}$ point we need to (i) approximate the spin-orbit interaction
$\tfrac{\hbar}{4m^2c^2}(\boldsymbol{\nabla}V\times\mathbf{p})\cdot\boldsymbol{s}$ with the bilayer graphene point symmetry by the atomic (and hence
isotropic) spin-orbit interaction $\xi\,\mathbf{L}\cdot\boldsymbol{s}$ and (ii) take into account the $d_{\pm}$ states in the effective $p_z^{\rm{eff}}$
orbitals, which enter the on-site Bloch wave functions, see Eqs.~(\ref{TB-basis}), (\ref{TB-basis-spin}) and (\ref{pz-eff}).
The on-site Bloch wave functions carrying the pseudospin A are formed by $d_{+}$ orbitals contrary the on-site wave functions labeled by the pseudospin B
which are composed of $d_{-}$ orbitals.
The atomic spin-orbit energies of $d_{+}\otimes\mid\ua\rangle$ and $d_{-}\otimes\mid\da\rangle$ are equal and higher by
$2\xi_d$ than the energies of $d_{+}\otimes\mid\da\rangle$ and $d_{-}\otimes\mid\ua\rangle$.
Hence we expect the general spectral tendency---the spin-orbit pulling the on-site Bloch states $\Psi_{\rm{A},\ua}$ and $\Psi_{\rm{B},\da}$ higher in energy
by $4\gamma^2\xi_d$ compared to $\Psi_{\rm{A},\da}$ and $\Psi_{\rm{B},\ua}$.

Let us firstly analyze the zero external electric field case without the spin-orbit interaction. When two single-layer graphene sheets
are brought together, the Bloch orbital states on the interlayer-direct-contact atoms ${\rm{A}}_1$ and ${\rm{B}}_2$ start to interact (via spin independent parameter $\gamma_1$)
and they form the high-energy antibonding (conduction) and bonding (valence) states $\Psi_{{\rm{A}}_1}+\Psi_{{\rm{B}}_2}$ and $\Psi_{{\rm{A}}_1}-\Psi_{{\rm{B}}_2}$,
respectively.
The low-energy states, on the other hand, are formed by the indirect-contact Bloch orbitals $\Psi_{{\rm{B}}_1}$ and $\Psi_{{\rm{A}}_2}$. If we now turn on
the spin-orbit interaction and count the spin degrees of freedom we would see the following changes in the bilayer energetics:
The two antibonding (bonding) states
$\Psi_{{\rm{A}}_1,\ua}+\Psi_{{\rm{B}}_2,\ua}$ and $\Psi_{{\rm{A}}_1,\da}+\Psi_{{\rm{B}}_2,\da}$
($\Psi_{{\rm{A}}_1,\ua}-\Psi_{{\rm{B}}_2,\ua}$ and $\Psi_{{\rm{A}}_1,\da}-\Psi_{{\rm{B}}_2,\da}$) should stay spin unsplit since the opposite pseudospin
components $\Psi_{{\rm{A}},s}$ and $\Psi_{{\rm{B}},s}$ entering the antibonding (bonding) wave functions are shifted opposite in energy by
$\xi\,\mathbf{L}\cdot\boldsymbol{s}$ and hence there is no net spin splitting. The situation is different for the four low-energy states $\Psi_{{\rm{B}}_1,s}$
and $\Psi_{{\rm{A}}_2,s}$. Their four-fold degeneracy is partially lifted when the spin-orbit interaction is turned on; states $\Psi_{{\rm{B}}_1,\da}$ and
$\Psi_{{\rm{A}}_2,\ua}$ remain degenerate and become shifted in energy higher than the degenerate pair $\Psi_{{\rm{B}}_1,\ua}$ and
$\Psi_{{\rm{A}}_2,\da}$. This splitting is seen in the spectrum as the $\mathrm{K}$ point anticrossing (see Fig.~\ref{bs0evpernm}), and according to
the above qualitative model we can fix the value of $4\gamma^2\xi_d$ to $24\,\rm{\mu eV}$. This reasoning is fully consistent with the tight-binding
energy spectrum when plugging in Eqs.~(\ref{spectrumatKone}-\ref{spectrumatKeight}) zero for $V$ and $\lambda_1$ and setting
$\lambda_{{\rm{I}}1}=\lambda_{{\rm{I}}1}'$. The spectral situation is schematically depicted on Fig.~\ref{energyspectrum}.

An external electric field breaks inversion symmetry and causes external spin splittings we observe in the first-principles spectra. Since the first layer is placed
to the potential $V$ and the second layer to the potential $-V$, we separate the spin-split states $\Psi_{{\rm{B}}_1,\da}$ and $\Psi_{{\rm{B}}_1,\ua}$ away from the
spin-split states $\Psi_{{\rm{A}}_2,\ua}$ and $\Psi_{{\rm{A}}_2,\da}$, what manifests in the low-energy spectrum as the band gap opening,
see Figs.~\ref{bs0.025evpernm}-\ref{bs10evpernm}. The energetics of the high-energy bands is somewhat different. The applied electric field affects
the antibonding (bonding) Bloch states by raising the relative magnitude of the pseudospin component $\Psi_{{\rm{A}}_1}$ ($\Psi_{{\rm{B}}_2}$) over
the component $\Psi_{{\rm{B}}_2}$ ($\Psi_{{\rm{A}}_1}$), see the behavior of $\alpha_{\pm}$ given by Eq.~(\ref{alpha}). The corresponding spin splitting is then dictated by the dominant pseudospin
orbital $\Psi_{{\rm{A}}_1}$ ($\Psi_{{\rm{B}}_2}$) when coupled to the spin, i.e. $\Psi_{{\rm{A}}_1,\ua}$ ($\Psi_{{\rm{B}}_2,\da}$) goes in energy higher
than $\Psi_{{\rm{A}}_1,\da}$ ($\Psi_{{\rm{B}}_2,\ua}$). This tendency is again well confirmed by our tight-binding model spectrum,
Eqs.~(\ref{spectrumatKone}-\ref{spectrumatKeight}), and its eigenstates distribution, Fig.~\ref{energyspectrum}, and as
well by the first-principles calculations.

\subsubsection{Interlayer spin-orbit couplings}

\begin{figure}
\includegraphics[width=0.8\columnwidth]{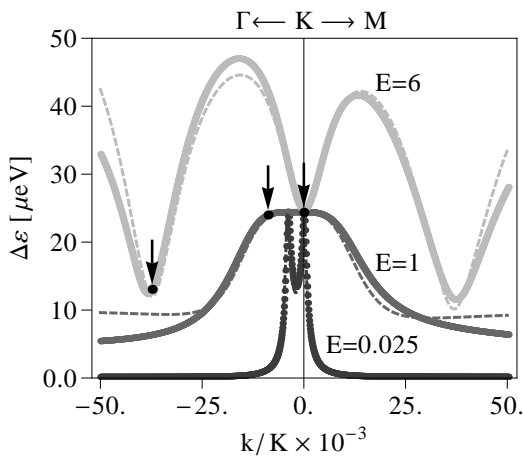}
\caption{Calculated spin splittings of the low-energy-conduction band for the electric fields of $E=25\,\rm{mV/nm}$,
$E=1\,\rm{V/nm}$, and $E=6\,\rm{V/nm}$. The solid lines are first-principles and dashed lines tight-binding results
with the intra-layer spin-orbit effects. The
arrows and circles indicate the positions of the conduction band minima.
}\label{LCB_splitting_asE}
\end{figure}

\begin{figure}
\includegraphics[width=0.8\columnwidth, angle=0]{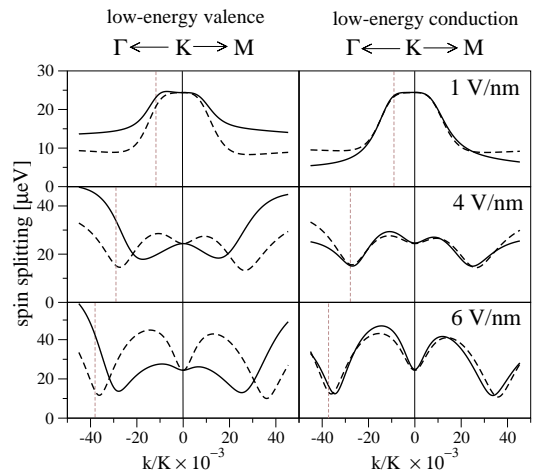}
\caption{Effects of the intra-layer spin-orbit couplings.
Calculated spin splittings of the low-energy valence (left) and
conduction (right) bands. First-principles results are shown as solid
lines, while the tight-binding fits with intra-layer spin-orbit parameters
only are dashed. The maxima of the valence bands and the minima of
the conduction bands are indicated by thin dashed vertical lines.
The conduction band spin physics appears satisfactorily
described, but the valence band splittings are rather off for the
tight-binding model.
}
\label{fits_low_intra}
\end{figure}

The intra-layer spin-orbit couplings $\lambda_{\rm{I}1}$, $\lambda_{\rm{I}2}$ and $\lambda_{\rm{BR}}$ suffice
to explain the bilayer spectrum
in the presence of a transverse electric field directly at the $\rm{K}$ points.
Around $\rm{K}$ points, including the regions of the valence band maxima and
the conduction band minima, the description is satisfactory for the
conduction band only, see Fig.~\ref{LCB_splitting_asE}. However, if
we look at the low-energy valence band (or high-energy bands), we see that the
fine features of the spin splittings differ from what the tight-binding
model with intra-layer spin-orbit coupling predicts. This
is markedly seen in Fig.~\ref{fits_low_intra}.

In the following we include into the picture
interlayer spin-orbit couplings, motivated by our symmetry-derived Hamiltonian
in App.~A, and demonstrate a very good quantitative agreement with
first-principles data. We stress that, (i) the interlayer spin-orbit
couplings are of the same order as the intra-layer ones, that is
typically $10 \mu$eV, and (ii) our fitting, while physically motivated
and robust, can be in principle non-unique as there are in principle 10 parameters entering
the spin-orbit Hamiltonian, see Eq.~(\ref{Heff_SOC}).
As such, the presented model, described by Hamiltonian (\ref{Heff_bi}), should be considered
as a physically reasonable convenient minimum quantitative description of the
spin-orbit physics in bilayer graphene with broken space inversion
symmetry by a transverse electric field. If one aims to describe the low-energy
conduction band only, one can neglect these interlayer spin-orbit couplings
entirely.

\begin{figure}
\includegraphics[width=0.8\columnwidth, angle=0]{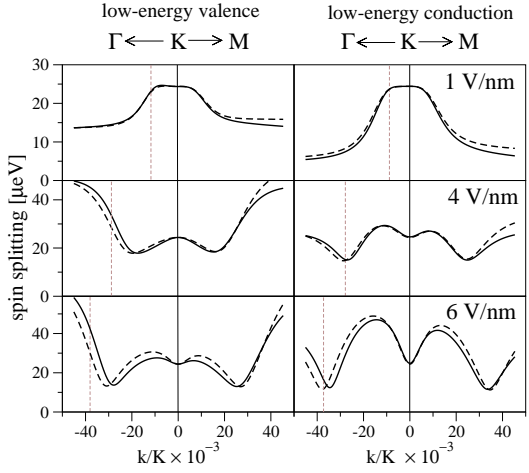}
\caption{Best fitting tight-binding model.
Calculated spin splittings of the low-energy valence (left) and
conduction (right) bands. First-principles results are shown as solid
lines, while the tight-binding fits, including all possible k-independent
spin-orbit parameters, are dashed. The maxima of the valence bands and the minima of
the conduction bands are indicated by thin dashed vertical lines.
}
\label{fits_low}
\end{figure}

We now explain the fitting procedure. The spin-orbit Hamiltonian in the external electric field, as presented
by Eq.~(\ref{Heff_SOC}), or Tab. \ref{external SOC} in App.~A, contains 10 real $\lambda$ parameters.
The diagonal ones, the intrinsic $\lambda_{\rm{I}}$,
determine the spin splittings at the $\rm{K}$ point. They can be easily
fixed, and we find that they do not depend on the electric field significantly
(within one $\rm{\mu eV}$). We then look at the spectral changes of the spin splittings
around $\rm{K}$ point as different off-diagonal parameters vary, and compare the trends
in the changes with the actual first-principles data. Moreover, we assume that $2\lambda_{\rm{BR}}$, $\lambda_1$,
$\lambda_3$ and $\delta\lambda_4$, which are absent in zero electric field, scale linearly with the intensity of
the applied electric field. One can see that
$\lambda_1 \approx 0$, as changes in this parameter distort the picture
away from the first-principles results. The direct interlayer coupling
is then largely spin-independent (governed solely by $\gamma_1$).

The parameters $\lambda_0=\overline{\lambda}_0+2\lambda_{\rm{BR}}$ and $\lambda_0'=\overline{\lambda}_0-2\lambda_{\rm{BR}}$
describe both the global and local breaking of space inversion symmetry. In the absence of an electric field,
$\lambda_0'= \lambda_0 = \overline{\lambda}_0$, since $2\lambda_{\rm{BR}}$ vanishes, see App.~A.
The parameter $\overline{\lambda}_0$ describes a local bulk-inversion-asymmetry physics: the electrons in one layer feel
an effective electric field due to the presence of the other layer. This field
gives rise to a ``local Dresselhaus" \cite{Fabian2007:APS} spin-orbit coupling.
Naturally, the field is opposite in the two layers so the net effect is zero, as there is no
global bulk inversion asymmetry. We can estimate $\overline{\lambda}_0$
along the following lines. The direct interlayer coupling is given
by the energy $\gamma_1 \approx 0.3$ eV. Since the distance between the
two layers is about 0.3 nm, the effective electric field felt by each
layer due to the presence of the other is about 1 V/nm. We know
from single-layer graphene that such a field gives the (real) Bychkov-Rashba
splitting of $2\lambda_{\rm BR} = 10\,\rm{\mu eV}$. This gives an order of magnitude
estimate $\overline{\lambda}_0 \approx 10\,\rm{\mu eV}$, which gives also a check on
how reasonable the actual fit is.

\begin{figure}
\includegraphics[width=0.8\columnwidth, angle=0]{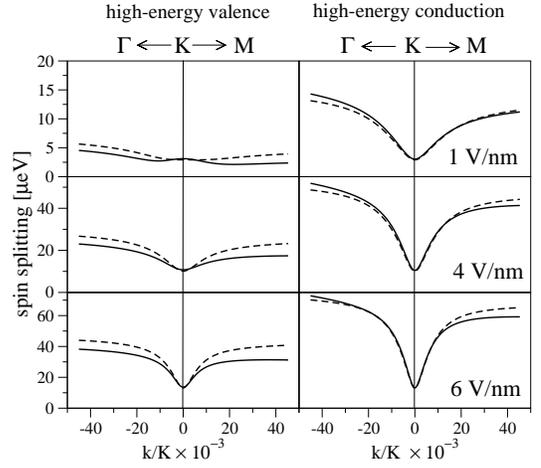}
\caption{Calculated spin splittings of the high-energy valence (left) and
conduction (right) bands. First-principles results are shown as solid
lines, while the tight-binding calculations, using the best fitting
parameters, are dashed.
}
\label{fits_high}
\end{figure}

In our fitting, knowing both the complexity and crudeness of the procedure,
we focused on obtaining both reasonable and robust results. To that end we
made a restricted least-squares fit to the first-principles spin-splittings
using the data for the electric fields of 1 and 6 V/nm. The fits were
simultaneous to both data sets, for the valence and conduction low-energy
bands only. The fitting was restricted to the spectrum around $\rm{K}$
within 2.5\% in the two directions $\rm{K}$-$\rm{\Gamma}$ and $\rm{K}$-$\rm{M}$, constraining
the fits to closely preserve the extremal points (minima and maxima) of the spectral
splittings, so that the overall shape was correct.
The crucial test of the robustness of the obtained parameters, and of
the assumption of the linearity (in $E$) of the spin-orbit
parameters, was (a) reproducing the spin splittings of the high-energy
bands which were not used in the fitting procedure, and (b), reproducing
the spin splittings of the low- and high-energy bands at intermediate
electric fields, which were also not included in the fitting. The actual
parameters of the fits are given as follows:
\begin{equation}
\begin{aligned}
\overline{\lambda}_0&=5\,\rm{\mu eV}\,, & 2\lambda_{\rm{BR}}&=\hspace{1mm}10\times E[{\rm V/nm}]\,\mu{\rm eV}\,,\\
\lambda_1&=0\,, &  \lambda_3&=1.5\times E[{\rm V/nm}]\,\mu{\rm eV}\,,\\
\overline{\lambda}_4&=-12\,\rm{\mu eV}\,, & \delta \lambda_4&=-3\times E[{\rm V/nm}]\,\mu{\rm eV}\,,
\end{aligned}
\end{equation}
where the numerical value of the electric field intensity $E$ should be taken in the units of ${\rm V/nm}$.

\begin{figure}
\includegraphics[width=0.9\columnwidth]{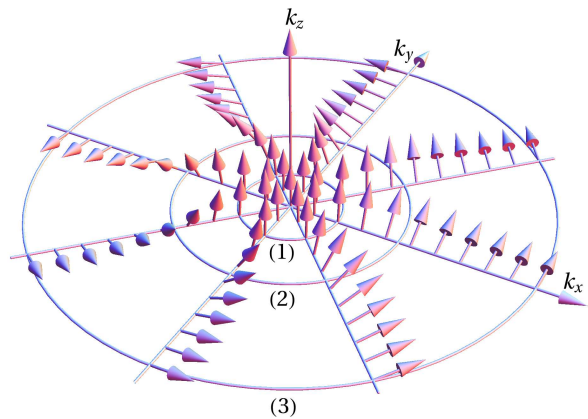}
\caption{(Color online) Calculated, from tight-binding, the spin orientation of the low-energy conduction-band states
as function of the momentum $\mathbf{k}$ for different directions for the electric field of $E=\rm 1\ V/nm$.
The $k_x$ axis corresponds to the $\rm{\Gamma KM}$ line. The circles corresponds
to (1) $|\mathbf{k}|/$\rm{K}$=0.01$, with the corresponding energy $\varepsilon=56\,\rm meV$ and the angle between the spin pointing vector
and the $k_z$ axis $\theta = 5.5^{\circ}$; (2) $|\mathbf{k}|/$\rm{K}$=0.0225$, $\varepsilon=119\,\rm  meV$, and $\theta = 45.5^{\circ}$; and
(3) $|\mathbf{k}|/$\rm{K}$=0.05$,  $\varepsilon=355\,\rm  meV$, and $\theta = 80.5^{\circ}$.
} \label{spinaligment}
\end{figure}

The first-principles data and the tight-binding fits for the low-energy bands
at electric fields of 1, 4, and 6\,$\rm{V/nm}$, are shown in Fig.~\ref{fits_low}.
The first-principles results for 4\,$\rm{V/nm}$ is, as discussed above, were not
used in the fitting of spin-orbit couplings (calculations for other fields give similar
level of agreement).
The fact that the first-principles results are well reproduced signifies
the robustness of the procedure and validity of our assumptions. The
spin splittings in bilayer graphene are rather complex, also considering
that the interesting points are not really $\rm{K}$ but the positions of the
valence band maxima and conduction bands minima. There is a clear
competition between the intrinsic splitting $\lambda_{\rm{I}}$, dominating close
to $\rm{K}$, and the extrinsic (off-diagonal) splittings, dominating at
momenta away from the $\rm{K}$ point. The spin pattern of the low energy conduction
band is shown in Fig.~\ref{spinaligment}, at the $\rm{K}$ point the spin quantization
axis is along $z$. This is due to the intrinsic spin-orbit coupling. Away from the
$\rm K$ point the spin quantization axis is in the plane, reflecting the dominance of
extrinsic SOC.

The splittings of the high-energy bands, which normally do not play a role
in transport, provide additional assurance in the
fitting. The results from our first-principles and tight-binding calculations
(with the parameters obtained by fitting the low-energy bands only) are shown in
Fig.~\ref{fits_high}. The quantitative agreement is very satisfactory.

The spin-orbit parameters entering the above fitting problem are in general momentum dependent,
see generalized $\mathbf{k}$-dependent spin-orbit Hamiltonian in Table \ref{external SOC-k-dependence}
of Appendix A.
However, our numerical analysis shows that for the momentum vectors within the considered 2.5\% interval around the $\rm{K}$ point
the values of the spin-orbit parameters modify less than 0.8\%, the spin-orbit Hamiltonian at the $\rm{K}$ point, Eq.~(\ref{Heff_SOC}),
is thus satisfactory.


\section{Conclusions}

We have systematically investigated the spin-orbit coupling effects in bilayer graphene, both intrinsic and
extrinsic, in the presence of a transverse electric field. We have presented first-principles results, and
analyzed them from a tight-binding perspective. We have derived and used the most general spin-orbit Hamiltonian for the
bilayer graphene with an absent space inversion (gated bilayer, bilayer on a substrate, or with adatoms) employing
the underlying bilayer graphene symmetries.
We have shown that the rough features of the spin splittings of the low-energy bands are well reproduced using a model with
single-layer-like spin-orbit couplings and interlayer orbital hoppings. In particular, the intrinsic anticrossings at zero
external electric field are fully consistent with the anticrossing mechanism proposed for the single-layer graphene.
In the presence of an electric field, the spin splittings of the (otherwise spin degenerate) bands are more subtle and complex
than in a single layer graphene. Directly at the $\rm{K}$ point the value of the splitting of the low-energy bands is given by
the intrinsic spin-orbit coupling. On the other hand, the high-energy bands are split in a proportion to the electric field, which
is what is normally expected.
Away from the $\rm K$ points, our spin-orbit upgraded tight-binding model gives an excellent description of the fine
spin splittings due to electric field. With the help of the interlayer spin-orbit couplings we have fully covered the first-principles
energetics at and near the $\rm{K}$ point.
A quantitative and physically inspired fitting procedure was proposed to obtain the realistic values of the interlayer
spin-orbit coupling parameters; we have shown that these parameters have similar values as the intra-layer ones.
The spin-orbit Hamiltonian we have proposed can be used in model studies of spin-dependent transport and spin relaxation in
extended and confined bilayer graphene.

\begin{acknowledgments}
We acknowledge support from the DFG SFB 689 and GRK 1459.
\end{acknowledgments}

\appendix
\section{Effective spin-orbit Hamiltonian at $\rm{K}(\rm{K}')$ from symmetry analysis}

The extrinsic spin-orbit coupling is induced by a transverse electric field which breaks the bilayer space inversion symmetry.
Using symmetry group arguments we express the corresponding spin-orbit Hamiltonian at $\rm K$ in the basis of the on-site Bloch wave functions
$\Psi_{{\rm A}_i,s}$ and $\Psi_{{\rm B}_i,s}$ [see Eqs.~(\ref{TB-basis-spin}) and (\ref{short-handed})], in the format of Tab. \ref{external SOC}.
This Hamiltonian matrix has 10 real parameters $\{\lambda_{{\rm I}1}, \lambda_{{\rm I}1}^\prime, \lambda_{{\rm I}2}, \lambda_{{\rm I}2}^\prime, \lambda_{0}=\overline{\lambda}_0+2\lambda_{\rm BR}, \lambda_{0}^\prime=\overline{\lambda}_0-2\lambda_{\rm BR},
\lambda_{4}=\overline{\lambda}_4+\delta\lambda_4, \lambda_{4}^\prime=\overline{\lambda}_4-\delta\lambda_4, \lambda_{1}, \lambda_{3}\}$.
The first four parameters are related to the intrinsic intra-layer spin-orbit coupling $\lambda_{\rm I}$ (hence subscript $\rm{I}$); the rest (extrinsic) are labeled corresponding
to the hopping parameters $\gamma_i$ with $i=\{0,1,3,4\}$.

\begin{table}[h!]
\begin{center}
\begin{tabular}{l|rccccccc}\hline\hline\\
${\rm SOC}$  & $\Psi_{{\rm A}_1,\ua}$ &  $\Psi_{{\rm A}_1,\da}$ &  $\Psi_{{\rm B}_1,\ua}$  &  $\Psi_{{\rm B}_1,\da}$    & $\Psi_{{\rm A}_2,\ua}$ &  $\Psi_{{\rm A}_2,\da}$ &  $\Psi_{{\rm B}_2,\ua}$  &  $\Psi_{{\rm B}_2,\da}$  \\\\\hline \\
\vspace*{-0.6em}
$\Psi_{{\rm A}_1,\ua}$ & $\lambda_{{\rm I}2}$  &                   0    &        0                &    0                      &    0                  & $\i\lambda_4$           & $\lambda_1$             & 0 \\ \\
\vspace*{-0.6em}
$\Psi_{{\rm A}_1,\da}$ &                    0  & $-\lambda_{{\rm I}2}$  & $\i\lambda_{0}$          &    0                      &  0                    &     0                  &                    0    & $-\lambda_1$ \\ \\
\vspace*{-0.6em}
$\Psi_{{\rm B}_1,\ua}$ &              0        & $-\i\lambda_0$          & $-\lambda_{{\rm I}1}$   & 0                         &   0                   &  0                     &                 0       & $-\i\lambda_4^\prime$ \\ \\
\vspace*{-0.6em}
$\Psi_{{\rm B}_1,\da}$ &         0             &        0               &    0                    & $\lambda_{{\rm I}1}$      & $\i\lambda_3$          &  0                     &               0         &  0 \\ \\
\vspace*{-0.6em}
$\Psi_{{\rm A}_2,\ua}$ &         0             &       0                &   0                     &       $-\i\lambda_3$       & $\lambda_{{\rm I}1}^\prime$ &    0             &                0        &  0 \\ \\
\vspace*{-0.6em}
$\Psi_{{\rm A}_2,\da}$ & $-\i\lambda_4$         &        0               &   0                     &  0                        &      0                      & $-\lambda_{{\rm I}1}^\prime$ & $-\i\lambda_{0}^\prime$ &  0 \\ \\
\vspace*{-0.6em}
$\Psi_{{\rm B}_2,\ua}$ & $\lambda_1$           &      0                 &       0                 &  0                        &  0                          & $\i\lambda_{0}^\prime$        & $-\lambda_{\rm{I}2}^\prime$ &  0 \\ \\
\vspace*{-0.6em}
$\Psi_{{\rm B}_2,\da}$ &            0          & $-\lambda_1$           & $\i\lambda_{4}^\prime$   &  0                        &    0                        &                0             &                           0 & $\lambda_{\rm{I}2}^\prime$   \\ \\
\vspace*{-0.6em}
&&&&&&&&\\ \hline\hline
\end{tabular}
\caption{Spin-orbit matrix elements in bilayer graphene with external electric field at $\rm K$.}\label{external SOC}
\end{center}
\end{table}

We now present the group theory analysis which leads to the spin-orbit coupling Hamiltonian
in Tab.~\ref{external SOC}. In the absence of an electric field the point group symmetry of the bilayer unit cell is $D_{3d}$, while the small
group of $\rm K$ is $D_3$. A transverse electric field along $z$ breaks the space inversion symmetry.
The point group reduces to $C_{3v}$ and the small group of $\rm K$ reduces to $C_3$.
The abelian group $C_3=\{E, R_{2\pi/3},R_{-2\pi/3}\}$ has three one-dimensional irreducible representations (we adopt
the notation from Ref.~[\onlinecite{KDWS1963:MIT}]), reproduced in Tab. \ref{character}. The complex conjugate representations
are related to the given ones by $\overline{\Gamma}_1=\Gamma_1$ and $\overline{\Gamma}_2=\Gamma_3$.

\begin{table}[h!]
\begin{minipage}[b]{0.43\linewidth}\centering
\begin{tabular}{l|ccc}
\hline\hline\\
\vspace*{-0.2em}
representation & \ \ $\Gamma_1$ & \ \ $\Gamma_2$ & $\Gamma_3$ \\ \\
\vspace*{-0.5em}
character $\chi_{{E}}$ & 1 & 1 & 1  \\ \\
\vspace*{-0.5em}
character $\chi_{R}$ & 1 & $\e^{\i\tfrac{2\pi}{3}}$ & $\e^{-\i\tfrac{2\pi}{3}}$ \\ \\
\vspace*{-0.5em}
character $\chi_{R^{-1}}$ & 1 & $\e^{-\i\tfrac{2\pi}{3}}$ & $\e^{\i\tfrac{2\pi}{3}}$ \\
&&&\\ \hline\hline
\end{tabular}
\end{minipage}
\hspace{1cm}
\begin{minipage}[b]{0.43\linewidth}\centering
\begin{tabular}{c|ccc}
\hline\hline\\
\vspace*{-0.5em}
$\Gamma\times\Gamma$\,  & \  $\Gamma_1$ &  \ $\Gamma_2$ & \ $\Gamma_3$ \\ \\
\hline\\
\vspace*{-0.6em}
$\Gamma_1$ & \ $\Gamma_1$ & \ $\Gamma_2$ & \ $\Gamma_3$ \\ \\
\vspace*{-0.5em}
$\Gamma_2$ & \ $\Gamma_2$ &  \ $\Gamma_3$ & \ $\Gamma_1$ \\ \\
\vspace*{-0.5em}
$\Gamma_3$ & \ $\Gamma_3$ & \ $\Gamma_1$ &\  $\Gamma_2$ \\
&&&\\ \hline\hline
\end{tabular}
\end{minipage}
\caption{Irreducible representations of $C_3$ and their characters (left) and the corresponding multiplication table (right)}\label{character}
\end{table}

The orbital components $\Psi_{{\rm A}_i}$ and $\Psi_{{\rm B}_i}$ of the on-site Bloch functions
$\Psi_{{\rm A}_i,s}$ and $\Psi_{{\rm B}_i,s}$ transform
according to $R_{2\pi/3}$, with the point-symmetry rotation axis given by $\rm{A}_1$-$\rm{B}_2$, as follows:
\begin{equation}
\begin{aligned}
R_{2\pi/3}(\Psi_{{\rm A}_1})&=\Psi_{{\rm A}_1},  &  R_{2\pi/3}(\Psi_{{\rm B}_1})&=\e^{\i\frac{2\pi}{3}}\,\Psi_{{\rm B}_1},\\
R_{2\pi/3}(\Psi_{{\rm A}_2})&=\e^{-\i\frac{2\pi}{3}}\,\Psi_{{\rm A}_2}, &  R_{2\pi/3}(\Psi_{{\rm B}_2})&=\Psi_{{\rm B}_2}.
\end{aligned}
\end{equation}
These transformations follow immediately from definition, Eq.~(\ref{TB-basis}), when substituting for $\mathbf{k}$-vector the $\rm{K}$ point momentum
and employing the $\pi$-state property of $p_z^{\rm{eff}}$ orbitals.
Hence $\Psi_{{\rm A}_1}$ and
$\Psi_{{\rm B}_2}$ belong to the trivial representation $\Gamma_1$, while $\Psi_{{\rm B}_1}\in\Gamma_2$ and
$\Psi_{{\rm A}_2}$ to $\Gamma_3$. Expressing the spin-orbit coupling in terms of $s_z$ and the spin raising and lowering matrices
$s_{\pm}=\tfrac{1}{2}(s_x \pm \i s_y)$,
\begin{equation}\label{SOC}
    H_{\rm SO}=\frac{\hbar}{4m^2c^2}\bigl(\boldsymbol{\nabla}V\times\mathbf{p}\bigr)\cdot\boldsymbol{s}\sim L_z s_z+L_{-} s_{+}+L_{+} s_{-}\,,
\end{equation}
we get operators $L_z$ and $L_\pm=L_x\pm \i L_y$ which act only on the orbital part of the on-site Bloch
wave functions $\Psi_{{\rm A}_i,s}$ and $\Psi_{{\rm B}_i,s}$.
The total potential $V=V_{\rm bi}+2V_{\rm el}$, entering Eq.~(\ref{SOC}),
comprises the intrinsic bilayer potential energy and the energy due to the externally applied
electric field. With respect to $C_3$ the $L$ operators transform as,
\begin{equation}\label{L-operators}
R_{2\pi/3} L_z= L_z R_{2\pi/3},\  R_{2\pi/3} L_{\pm}= \e^{\mp \i\frac{2\pi}{3}}\,L_{\pm} R_{2\pi/3}.
\end{equation}
Hence, $L_z\in\Gamma_1$, $L_{-}\in\Gamma_2$, and $L_{+}\in\Gamma_3$. The matrix element $\langle \Psi|L|\Phi\rangle$ belongs to the
$\overline{\Gamma}_\Psi\times\Gamma_L\times\Gamma_\Phi$ representation of the group $C_3$. According to the (abelian) group theory:
$$
\mbox{if}\ \ \overline{\Gamma}_\Psi\times\Gamma_L\times\Gamma_\Phi\neq\Gamma_1\,,\ \ \mbox{than}\ \ \langle \Psi|L|\Phi\rangle=0\,.
$$
Using this fact and the multiplication Tab.~\ref{character}, one can identify all the vanishing Hamiltonian matrix
elements in Tab.~\ref{external SOC}. It is also clear that
all diagonal matrix elements are real. We show below that the off-diagonal elements are either real or pure imaginary.

The small group $C_3$ of $\rm K$ is a subgroup of the point group $C_{3v}$ which in addition to the two rotations
$\{R_{\pm 2\pi/3}\}$ also has three vertical reflections
$\{R_v, R_{\pm 2\pi/3}\circ R_v\}$. Here $R_v$ is the
reflection with respect to the vertical plane defined by the electric field ($z$-axis) and the line connecting atoms
${\rm A}_1$ and ${\rm B}_1$, see Fig.~(\ref{BiStructure}) for the coordinate system we use.
Expressing the action of $R_v$ in coordinates,
$$
R_v:\mathbf{r}=(x,y,z)\mapsto(R_v\mathbf{r})=(-x,y,z)\,,
$$
we see that this transformation preserves the layer label and the pseudospin (${\rm{A}}_i\leftrightarrow{\rm{A}}_i$ and ${\rm{B}}_i\leftrightarrow{\rm{B}}_i$),
but sends $\rm K$ to $\rm{K}'$. The action of
$R_v$ on the on-site space Bloch wave functions and $L$ operators is therefore as follows:
\begin{equation}
\begin{aligned}
R_v(\Psi_{{\rm A}_i,\mathbf{K}})&=\Psi_{{\rm A}_i,\mathbf{K}^\prime}=\Psi_{{\rm A}_i,\mathbf{K}}^*\,, \ &\ R_v L_z&=-L_zR_v\\
R_v(\Psi_{{\rm B}_i,\mathbf{K}})&=\Psi_{{\rm B}_i,\mathbf{K}^\prime}=\Psi_{{\rm B}_i,\mathbf{K}}^*\,, \ &\ R_v L_{\pm}&=L_{\mp}R_v\,.
\end{aligned}
\end{equation}
Moreover,
$$
\begin{aligned}
\langle\Psi\,|\,\Phi\rangle&=\int d^3\mathbf{r}\, \Psi^*(\mathbf{r})\,\Phi(\mathbf{r})\\
&=\int d^3\mathbf{r}\, \Psi^*(R_v\mathbf{r})\,\Phi(R_v\mathbf{r})=\langle R_v(\Psi)\,|\,R_v(\Phi)\rangle\,,
\end{aligned}
$$
that is, $R_v$ is a unitary operator when acting on wave functions.
The above facts suffice to show that, for example, $\langle \Psi_{{\rm A}_1}|L_z|\Psi_{{\rm B}_2}\rangle$ is real. Indeed,
$$
\begin{aligned}
\langle \Psi_{{\rm A}_1,\mathbf{K}}\,|\,L_z \Psi_{{\rm B}_2,\mathbf{K}}\rangle&=\langle (L_z\Psi_{{\rm B}_2,\mathbf{K}})^*\,|\,(\Psi_{{\rm A}_1,\mathbf{K}})^*\rangle\\
&=\langle -L_z(\Psi_{{\rm B}_2,\mathbf{K}})^*\,|\,(\Psi_{{\rm A}_1,\mathbf{K}})^*\rangle\\
&=-\langle L_z\Psi_{{\rm B}_2,\mathbf{K}^\prime}\,|\,\Psi_{{\rm A}_1,\mathbf{K}^\prime}\rangle\\
&=-\langle L_z R_v(\Psi_{{\rm B}_2,\mathbf{K}})\,|\,R_v(\Psi_{{\rm A}_1,\mathbf{K}})\rangle\\
&=-\langle -R_v L_z(\Psi_{{\rm B}_2,\mathbf{K}})\,|\,R_v(\Psi_{{\rm A}_1,\mathbf{K}})\rangle\\
&=\langle L_z\Psi_{{\rm B}_2,\mathbf{K}}\,|\,\Psi_{{\rm A}_1,\mathbf{K}}\rangle\\
&=\langle \Psi_{{\rm A}_1,\mathbf{K}}\,|\,L_z \Psi_{{\rm B}_2,\mathbf{K}}\rangle^*\,.
\end{aligned}
$$
In a similar way one can show that the matrix elements, which comprise $L_{\pm}$ operators are imaginary. The only changes in the above computation
appear in the second and the fourth lines. In the second line one uses $(L_{\pm})^*=-L_{\mp}$ and in the fourth line the commutation relation
$L_{\mp}R_v=R_v L_{\pm}$ instead $L_zR_v=-R_v L_z$.

It is convenient to express the spin-orbit Hamiltonian given by Tab.~\ref{external SOC} in a more conventional form,
\begin{equation}\label{sum hamiltonian}
H_{\rm SO}=H_{{\rm I}}+H_{{\rm BR}}+H_{{\rm inter}}+H_{{\rm{el}}}\,.
\end{equation}
Here $H_{{\rm I}}$ and $H_{{\rm BR}}$ are the intrinsic and Bychkov-Rashba-like Hamiltonians, respectively.
Hamiltonian $H_{{\rm inter}}$ comprises terms
that are specific for interlayer coupling (bilayer geometry), and $H_{{\rm{el}}}$ contains
the remaining interlayer contributions present only in a finite transverse electric field.
We introduce the Pauli matrices $\boldsymbol{\mu}$ (layer), $\boldsymbol{\sigma}$ (sublattice pseudospin), $\boldsymbol{s}$ (spin), along
with the unit $2\times 2$ matrices $\mu_0$, $\sigma_0$, $s_0$, and symmetrized parameters:
\begin{equation}\label{symmetrized1}
\begin{aligned}
\Delta_{{\rm I}}&=\tfrac{1}{2}(\lambda_{{\rm I}2}-\lambda_{{\rm I}1}) \ \  &\ \  \Delta_{{\rm I}}^\prime&=\tfrac{1}{2}(\lambda_{{\rm I}2}^\prime-\lambda_{{\rm I}1}^\prime)\\
\lambda_{{\rm I}}&=\tfrac{1}{2}(\lambda_{{\rm I}2}+\lambda_{{\rm I}1}) \ \  &\ \  \lambda_{{\rm I}}^\prime& =\tfrac{1}{2}(\lambda_{{\rm I}2}^\prime+\lambda_{{\rm I}1}^\prime)
\end{aligned}
\end{equation}
and
\begin{equation}\label{symmetrized2}
\begin{aligned}
2\lambda_{\rm BR}&=\tfrac{1}{2}(\lambda_{0}-\lambda_{0}^\prime)  \ \  &\ \ \delta\lambda_{4}&=\tfrac{1}{2}(\lambda_{4}-\lambda_{4}^\prime)\\
\overline{\lambda}_{0}&=\tfrac{1}{2}(\lambda_{0}+\lambda_{0}^\prime)  \ \  &\ \ \overline{\lambda}_{4}&=\tfrac{1}{2}(\lambda_{4}+\lambda_{4}^\prime)\,.
\end{aligned}
\end{equation}
We get the following effective Hamiltonian:
\begin{equation}
\begin{aligned}
H_{{\rm I}}=\,&\tfrac{1}{2}\bigl[(\Delta_{{\rm I}}-\Delta_{{\rm I}}^\prime)\sigma_0+(\lambda_{{\rm I}}+\lambda_{{\rm I}}^\prime)\sigma_z\bigr]\mu_0 \tau s_z\\
+&\tfrac{1}{2}\bigl[(\Delta_{{\rm I}}+\Delta_{{\rm I}}^\prime)\sigma_0+(\lambda_{{\rm I}}-\lambda_{{\rm I}}^\prime)\sigma_z\bigr]\mu_z \tau s_z\,,
\end{aligned}
\end{equation}

\begin{equation}
\begin{aligned}
H_{{\rm BR}}&=\tfrac{1}{2}(\overline{\lambda}_{0}\mu_z+2\lambda_{\rm BR}\mu_0)(\tau \sigma_x s_y-\sigma_y s_x)\,,\\
\end{aligned}
\end{equation}

\begin{equation}
\begin{aligned}
H_{{\rm inter}}&=-\tfrac{1}{2}(\overline{\lambda}_{4}\sigma_z+\delta\lambda_{4}\sigma_0)(\tau \mu_x s_y+\mu_y s_x)\,,\\
\end{aligned}
\end{equation}

\begin{equation}
\begin{aligned}
&H_{{\rm el}}=\tfrac{\lambda_1}{2}\tau s_z(\mu_x\sigma_x-\mu_y
\sigma_y)\\
&+\tfrac{\lambda_3}{4}[\mu_x(\tau \sigma_x s_y+\sigma_y s_x)+\mu_y(\tau \sigma_y s_y-\sigma_x s_x)]\,.
\end{aligned}
\end{equation}
We have included $\tau = 1$ ($\rm K$) and $\tau = -1$ ($\rm{K}'$), to cover also $\rm{K}'$. The translation of the
Hamiltonian matrix elements from $\rm K$ to $\rm{K}'$ is based on time reversal symmetry.

In the absence of an electric field, the spin-orbit Hamiltonian of bilayer graphene was derived
in Ref.~[\onlinecite{Guinea2010:arxiv}]. This is a special case of our Hamiltonian
in Tab.~\ref{external SOC}. We can restore space inversion symmetry
\begin{equation}
\mathfrak{i}:\mathbf{r}=(x,y,z)\mapsto\mathfrak{i}(\mathbf{r})=(-x,-y,-z),
\end{equation}
mapping $\rm K$ to $\rm{K}'$ and interchange the layer indexes
and the pseudospin,
$$
{\rm{A}_1}\leftrightarrow{\rm{B}_2}\ \ \ {\rm{B}_1}\leftrightarrow{\rm{A}_2}\,.
$$
The orbital momentum $L$ operators are invariant with respect to inversion, which is unitary, that is,
 $\langle \Psi\,|\,\Phi\rangle=\langle \mathfrak{i}(\Psi)\,|\,\mathfrak{i}(\Phi)\rangle$.
The action of $\mathfrak{i}$ on the on-site space Bloch wave function is:
\begin{equation}
\begin{aligned}
\mathfrak{i}(\Psi_{{\rm A}_1,\mathbf{K}})&=\Psi_{{\rm B}_2,\mathbf{K}^\prime}=\Psi_{{\rm B}_2,\mathbf{K}}^*\,,\\
\mathfrak{i}(\Psi_{{\rm B}_2,\mathbf{K}})&=\Psi_{{\rm A}_1,\mathbf{K}^\prime}=\Psi_{{\rm A}_1,\mathbf{K}}^*\,,\\
\mathfrak{i}(\Psi_{{\rm B}_1,\mathbf{K}})&=\Psi_{{\rm A}_2,\mathbf{K}^\prime}=\Psi_{{\rm A}_2,\mathbf{K}}^*\,,\\
\mathfrak{i}(\Psi_{{\rm A}_2,\mathbf{K}})&=\Psi_{{\rm B}_1,\mathbf{K}^\prime}=\Psi_{{\rm B}_1,\mathbf{K}}^*\,.
\end{aligned}
\end{equation}
Applying the above facts to the nonzero matrix elements in Tab.~\ref{external SOC} we get the spin-orbit Hamiltonian of
Ref.~[\onlinecite{Guinea2010:arxiv}] reducing the number of free (real) parameters to four,
$\{\lambda_{{\rm I}1}=\lambda_{{\rm I}1}^\prime, \lambda_{{\rm I}2}=\lambda_{{\rm I}2}^\prime, \lambda_{0}=\lambda_{0}^\prime,
\lambda_{4}=\lambda_{4}^\prime, \lambda_1=\lambda_3 = 0\}$, given in Tab. \ref{guinea SOC}.

\begin{table}[h!]
\begin{center}
\begin{tabular}{l|rccccccc}\hline\hline\\
${\rm SOC}$  & $\Psi_{{\rm A}_1,\ua}$ &  $\Psi_{{\rm A}_1,\da}$ &  $\Psi_{{\rm B}_1,\ua}$  &  $\Psi_{{\rm B}_1,\da}$    & $\Psi_{{\rm A}_2,\ua}$ &  $\Psi_{{\rm A}_2,\da}$ &  $\Psi_{{\rm B}_2,\ua}$  &  $\Psi_{{\rm B}_2,\da}$  \\\\\hline \\
\vspace*{-0.6em}
$\Psi_{{\rm A}_1,\ua}$ & $\lambda_{{\rm I}2}$  &                   0    &        0                &    0                      &    0                  & $\i\lambda_4$           & 0            & 0 \\ \\
\vspace*{-0.6em}
$\Psi_{{\rm A}_1,\da}$ &                    0  & $-\lambda_{{\rm I}2}$  & $\i\lambda_{0}$          &    0                      &  0                    &     0                  &                    0    & 0 \\ \\
\vspace*{-0.6em}
$\Psi_{{\rm B}_1,\ua}$ &              0        & $-\i\lambda_0$          & $-\lambda_{{\rm I}1}$   & 0                         &   0                   &  0                     &                 0       & $-\i\lambda_4$ \\ \\
\vspace*{-0.6em}
$\Psi_{{\rm B}_1,\da}$ &         0             &        0               &    0                    & $\lambda_{{\rm I}1}$      & 0         &  0                     &               0         &  0 \\ \\
\vspace*{-0.6em}
$\Psi_{{\rm A}_2,\ua}$ &         0             &       0                &   0                     &       0       & $\lambda_{{\rm I}1}$ &    0             &                0        &  0 \\ \\
\vspace*{-0.6em}
$\Psi_{{\rm A}_2,\da}$ & $-\i\lambda_4$         &        0               &   0                     &  0                        &      0                      & $-\lambda_{{\rm I}1}$ & $-\i\lambda_{0}$ &  0 \\ \\
\vspace*{-0.6em}
$\Psi_{{\rm B}_2,\ua}$ & 0           &      0                 &       0                 &  0                        &  0                          & $\i\lambda_{0}$        & $-\lambda_{\rm{I}2}$ &  0 \\ \\
\vspace*{-0.6em}
$\Psi_{{\rm B}_2,\da}$ &            0          & 0           & $\i\lambda_{4}$   &  0                        &    0                        &                0             &                           0 & $\lambda_{\rm{I}2}$   \\ \\
\vspace*{-0.6em}
&&&&&&&&\\ \hline\hline
\end{tabular}
\caption{Spin-orbit Hamiltonian matrix elements in bilayer graphene in the absence of an external electric field at $\rm K$.}\label{guinea SOC}
\end{center}
\end{table}

We show, for example, that $\lambda_3=0$; by similar considerations one can check other spin-orbit elements:
$$
\begin{aligned}
\langle \Psi_{{\rm B}_1,\mathbf{K}}\,|\,L_{+} \Psi_{{\rm A}_2,\mathbf{K}}\rangle&=\langle (L_{+}\Psi_{{\rm A}_2,\mathbf{K}})^*\,|\,(\Psi_{{\rm B}_1,\mathbf{K}})^*\rangle\\
&=\langle -L_{-}(\Psi_{{\rm A}_2,\mathbf{K}})^*\,|\,(\Psi_{{\rm B}_1,\mathbf{K}})^*\rangle\\
&=-\langle L_{-}\Psi_{{\rm A}_2,\mathbf{K}^\prime}\,|\,\Psi_{{\rm B}_1,\mathbf{K}^\prime}\rangle\\
&=-\langle L_{-}\mathfrak{i}(\Psi_{{\rm B}_1,\mathbf{K}})\,|\,\mathfrak{i}(\Psi_{{\rm A}_2,\mathbf{K}})\rangle\\
&=-\langle \mathfrak{i}L_{-}(\Psi_{{\rm B}_1,\mathbf{K}})\,|\,\mathfrak{i}(\Psi_{{\rm A}_2,\mathbf{K}})\rangle\\
&=-\langle L_{-}\Psi_{{\rm B}_1,\mathbf{K}}\,|\,\Psi_{{\rm A}_2,\mathbf{K}}\rangle\\
&=-\langle \Psi_{{\rm B}_1,\mathbf{K}}\,|\,L_{-}^\dag \Psi_{{\rm A}_2,\mathbf{K}}\rangle\\
&=-\langle \Psi_{{\rm B}_1,\mathbf{K}}\,|\,L_{+} \Psi_{{\rm A}_2,\mathbf{K}}\rangle=0\,.
\end{aligned}
$$

Rewriting the matrix in Tab. \ref{guinea SOC} in terms of the symmetrized parameters as defined by Eqs.~(\ref{symmetrized1}) and~(\ref{symmetrized2}) we get the
spin-orbit coupling Hamiltonian in the absence of electric field (that is in the presence of space inversion symmetry)
in a more conventional notation: \cite{Guinea2010:arxiv}
\begin{equation}
\begin{aligned}
&H_{\rm SO}(E=0)= \lambda_{{\rm I}\,}\tau\mu_0\sigma_z s_z+\Delta_{{\rm I}\,}\tau\mu_z\sigma_0 s_z\\
&+\tfrac{\overline{\lambda}_{0}}{2}\mu_z(\tau\sigma_x s_y-\sigma_y s_x)- \tfrac{\overline{\lambda}_{4}}{2}\sigma_z(\tau\mu_x s_y+\mu_y s_x)\,.
\end{aligned}
\end{equation}

To explain quantitatively spin splittings of the low-energy conduction band, two spin-orbit parameters are important:
$\lambda_{\rm I}$ and $4 \lambda_{\rm BR} = (\lambda_0 - \lambda_0')$.
The interlayer spin-flip parameters $\lambda_1$, $\lambda_3$, $\overline{\lambda}_4$ and $\delta\lambda_4$ are important to
quantitatively fit the valence band. Our spin-orbit Hamiltonian is valid in a general case of
bilayer graphene with absent space inversion symmetry, such as gated bilayer or bilayer on a substrate. It
is likely that a stronger bonding substrate would make several of the ten parameters much larger than
they are in an applied field.

The spin-orbit Hamiltonian $H_{\rm{SO}}$ as presented in Table \ref{external SOC} is, strictly speaking, valid directly at the $\rm{K}(\rm{K}')$ point.
It can be extended to be $\mathbf{k}$-dependent.
The strategy employs: the on-site Bloch states for a general $\mathbf{k}$ vector, Eqs.~(\ref{TB-basis}), the $\pi$-state symmetry of $p_z^{\rm{eff}}$
orbitals, transformational properties of the spin-orbit $L$ operators, Eqs.~(\ref{L-operators}), time-reversal symmetry, and finally, the nearest neighbor (nn)
and the next-nearest neighbor (nnn) approximations, according to which:
\beq\label{k-dependence}
\begin{aligned}
&\langle\Psi_{{\rm{X}}_i}(\mathbf{k})|\,L\,|\Psi_{{\rm{Y}}_j}(\mathbf{k})\rangle\,\approx\,\e^{\i\mathbf{k}(\mathbf{t}_{{\rm{Y}}_j}-\mathbf{t}_{{\rm{X}}_i})}\,\times\\
\\
\vspace{-8mm}
&\times\,\sum\limits_{\rm{nn(n)}}\e^{\i\mathbf{k}\mathbf{R}_{\rm{nn(n)}}}\bigl\langle p_z^{\rm{eff}}({\rm{X}}_i)|\,L\,|p_z^{\rm{eff}}({\rm{Y}}_j+\mathbf{R}_{\rm{nn(n)}})\bigr\rangle\,.
\end{aligned}
\eeq
In the expression above ${\rm{X}}_i$ and ${\rm{Y}}_j$ stand for an arbitrary couple of atoms $\rm{A}_1$, $\rm{B}_1$, $\rm{A}_2$, $\rm{B}_2$
in the bilayer elementary cell. The underlying symmetries of bilayer graphene enable us to express:
$$
\langle p_z^{\rm{eff}}({\rm{X}}_i)|L|p_z^{\rm{eff}}({\rm{Y}}_j+\mathbf{R}_{\rm{nn(n)}})\rangle
=\e^{\pm\i\rm{\Phi}}\,\langle p_z^{\rm{eff}}({\rm{X}}_i)|L|p_z^{\rm{eff}}({\rm{Y}}_j)\rangle,
$$
where the phase factor is either $0$, $\tfrac{\pi}{3}$, or $\tfrac{2\pi}{3}$ depending on the atoms ${\rm{X}}_i$ and ${\rm{Y}}_j$.
This observation simplifies summation over $\mathbf{R}_{\rm{nn(n)}}$ in Eq.~(\ref{k-dependence}). When proceeding as explained
we arrive at the $\mathbf{k}$-dependent spin-orbit Hamiltonian shown in the Table \ref{external SOC-k-dependence}.
\begin{table*}
\begin{center}
\begin{tabular}{l|cccccccc}\hline\hline\\
${\rm SOC}\,(\mathbf{k})$            & $\Psi_{{\rm A}_1,\ua}(\mathbf{k})$                  & $\Psi_{{\rm A}_1,\da}(\mathbf{k})$                    & $\Psi_{{\rm B}_1,\ua}(\mathbf{k})$      & $\Psi_{{\rm B}_1,\da}(\mathbf{k})$     & $\Psi_{{\rm A}_2,\ua}(\mathbf{k})$         & $\Psi_{{\rm A}_2,\da}(\mathbf{k})$          & $\Psi_{{\rm B}_2,\ua}(\mathbf{k})$             &  $\Psi_{{\rm B}_2,\da}(\mathbf{k})$  \\\\\hline \\
\vspace*{-0.1em}
$\Psi_{{\rm A}_1,\ua}(\mathbf{k})$ & $\lambda_{{\rm I}2}\,{\rm{u}}(\mathbf{k})$          & $\i\mu_{\rm{A}_1}{\rm{v}}(\mathbf{k})$                  &        0                                & $\i\lambda_{0}\,{\rm{w}}^{*}(-\mathbf{k})$        &  0                                         & $\i\lambda_4\, {\rm{w}}^{*}(\mathbf{k})$               & $\lambda_1\,{\rm{u}}(\mathbf{k})$ & $\i\mu{\rm{v}}(\mathbf{k})$ \\ \\
\vspace*{-0.1em}
$\Psi_{{\rm A}_1,\da}(\mathbf{k})$ & $-\i\mu_{\rm{A}_1}{\rm{v}}^*(\mathbf{k})$              & $-\lambda_{{\rm I}2}\,{\rm{u}}(\mathbf{k})$           & $\i\lambda_{0}\,{\rm{w}}(\mathbf{k})$      &    0                                   &  $\i\lambda_4\, {\rm{w}}(-\mathbf{k})$        & 0                                           & $-\i\mu{\rm{v}}^{*}(-\mathbf{k})$                                              & $-\lambda_1\,{\rm{u}}(\mathbf{k})$ \\ \\
\vspace*{-0.1em}
$\Psi_{{\rm B}_1,\ua}(\mathbf{k})$ &              0                                      & $-\i\lambda_0\,{\rm{w}}^{*}(\mathbf{k})$                  & $-\lambda_{{\rm I}1}\,{\rm{u}}(\mathbf{k})$    & $-\i\mu_{\rm{B}_1}{\rm{v}}(\mathbf{k})$        &  0                                         & $\i\lambda_3 \,{\rm{z}}^{*}(-\mathbf{k})$              & 0                                              & $-\i\lambda_4^\prime\, {\rm{w}}^{*}(\mathbf{k})$ \\ \\
\vspace*{-0.1em}
$\Psi_{{\rm B}_1,\da}(\mathbf{k})$ & $-\i\lambda_{0}\,{\rm{w}}(-\mathbf{k})$         &        0                                              &   $\i\mu_{\rm{B}_1}{\rm{v}}^*(\mathbf{k})$     & $\lambda_{{\rm I}1}\,{\rm{u}}(\mathbf{k})$    & $\i\lambda_3\, {\rm{z}}(\mathbf{k})$          & 0                                           & $-\i\lambda_4^\prime\, {\rm{w}}(-\mathbf{k})$     &  0 \\ \\
\vspace*{-0.1em}
$\Psi_{{\rm A}_2,\ua}(\mathbf{k})$ &         0                                           & $-\i\lambda_4\, {\rm{w}}^{*}(-\mathbf{k})$                &   0                                     & $-\i\lambda_3\,{\rm{z}}^{*}(\mathbf{k})$          & $\lambda_{{\rm I}1}^\prime\,{\rm{u}}(\mathbf{k})$ & $-\i\mu_{{\rm{A}}_2}{\rm{v}}(\mathbf{k})$             & 0                                              &  $-\i\lambda_{0}^\prime\,{\rm{w}}^{*}(-\mathbf{k})$ \\ \\
\vspace*{-0.1em}
$\Psi_{{\rm A}_2,\da}(\mathbf{k})$ & $-\i\lambda_4\, {\rm{w}}(\mathbf{k})$           &        0                                              & $-\i\lambda_3\,{\rm{z}}(-\mathbf{k})$        &  0                                     &  $\i\mu_{{\rm{A}}_2}{\rm{v}}^*(\mathbf{k})$           & $-\lambda_{{\rm I}1}^\prime\,{\rm{u}}(\mathbf{k})$ & $-\i\lambda_{0}^\prime\,{\rm{w}}(\mathbf{k})$     &  0 \\ \\
\vspace*{-0.1em}
$\Psi_{{\rm B}_2,\ua}(\mathbf{k})$ & $\lambda_1\,{\rm{u}}(\mathbf{k})$ &      $\i\mu{\rm{v}}(-\mathbf{k})$                                              &       0                                 & $\i\lambda_{4}^\prime\,{\rm{w}}^{*}(-\mathbf{k})$ &  0                                         & $\i\lambda_{0}^\prime\,{\rm{w}}^{*}(\mathbf{k})$       & $-\lambda_{\rm{I}2}^\prime\,{\rm{u}}(\mathbf{k})$     &  $\i\mu_{{\rm{B}}_2}{\rm{v}}(\mathbf{k})$ \\ \\
\vspace*{-0.1em}
$\Psi_{{\rm B}_2,\da}(\mathbf{k})$ &            $-\i\mu{\rm{v}}^{*}(\mathbf{k})$                                       & $-\lambda_1\,{\rm{u}}(\mathbf{k})$  & $\i\lambda_{4}^\prime\,{\rm{w}}(\mathbf{k})$ &  0                                     & $\i\lambda_{0}^\prime\,{\rm{w}}(-\mathbf{k})$ & 0                                           & $-\i\mu_{{\rm{B}}_2}{\rm{v}}^*(\mathbf{k})$              & $\lambda_{\rm{I}2}^\prime\,{\rm{u}}(\mathbf{k})$  \\ \\
\vspace*{-0.6em}
&&&&&&&&\\ \hline\hline
\end{tabular}
\caption{$\mathbf{k}$-dependent spin-orbit matrix elements of the bilayer graphene Hamiltonian with external electric field within the nearest and next nearest
neighbor approximation. Apart from the previously discussed spin-orbit parameters, there appear new five real couplings: $\mu$, $\mu_{{\rm{A}_1}}$, $\mu_{{\rm{B}_1}}$,
$\mu_{{\rm{A}_2}}$, and $\mu_{{\rm{B}_2}}$, they are not present at the $\rm{K}(\rm{K}')$ point.}\label{external SOC-k-dependence}
\end{center}
\end{table*}

The structural spin-orbit functions $\rm{u}(\mathbf{k})$, $\rm{v}(\mathbf{k})$, $\rm{w}(\mathbf{k})$ and $\rm{z}(\mathbf{k})$ of the bilayer graphene are
\begin{widetext}
\begin{eqnarray}
{\rm{u}}(\mathbf{k}) &=&
-\frac{2}{3\sqrt{3}}\bigl[\sin{\mathbf{k}\cdot\mathbf{R}_1}+\sin{\mathbf{k}\cdot\mathbf{R}_2}+\sin{\mathbf{k}\cdot\mathbf{R}_3}\bigr]\,,\\
{\rm{v}}(\mathbf{k}) &=&
\frac{2}{3\sqrt{3}}\bigl[\sin{\mathbf{k}\cdot\mathbf{R}_1}+\e^{\i\tfrac{2\pi}{3}}\sin{\mathbf{k}\cdot\mathbf{R}_2}+\e^{-\i\tfrac{2\pi}{3}}\sin{\mathbf{k}\cdot\mathbf{R}_3}\bigr]\,,\\
{\rm{w}}(\mathbf{k}) &=&
\frac{1}{3}\,\e^{\i\mathbf{k}\cdot\mathbf{t}_{{\rm{B}}_1}}\bigl[1+\e^{\i\tfrac{2\pi}{3}+\i\mathbf{k}\cdot\mathbf{R}_2}+\e^{-\i\tfrac{2\pi}{3}-\i\mathbf{k}\cdot\mathbf{R}_3}\bigr]\,,\\
{\rm{z}}(\mathbf{k}) &=&
\frac{1}{3}\,\e^{\i\mathbf{k}\cdot(\mathbf{t}_{{\rm{A}}_2}-\mathbf{t}_{{\rm{B}}_1})-\i\mathbf{k}\cdot\mathbf{R}_2}\bigl[\e^{\i\mathbf{k}\cdot\mathbf{R}_3}+\e^{-\i\tfrac{2\pi}{3}}+\e^{\i\tfrac{2\pi}{3}-\i\mathbf{k}\cdot\mathbf{R}_1}\bigr]\,,
\end{eqnarray}
where $\mathbf{R}_1=a(1,0)$, $\mathbf{R}_2=\tfrac{a}{2}(-1,-\sqrt{3})$ and $\mathbf{R}_3=\tfrac{a}{2}(-1,\sqrt{3})$ are Bravais hexagonal lattice vectors
with the lattice constant $a=2.46${\AA}.

The full tight-binding Hamiltonian with spin-orbit terms can be folded down to an effective Hamiltonian for the low-energy conduction
and valence bands only. To this end we perform L\"owdin transformation \cite{Lowdin1965:PR,Konschuh2010:PRB} by projecting Hamiltonian
$H_{\rm{eff}}(\mathbf{k})$ given by Eq.~(\ref{Heff_bi}), into the subspace of low-energy states with the on-site Bloch basis
$\Psi_{\rm{B}_1,\ua}$, $\Psi_{\rm{B}_1,\da}$, $\Psi_{\rm{A}_2,\ua}$, and $\Psi_{\rm{A}_2,\da}$. We keep only the intra-layer spin-orbit
coupling parameters $\lambda_{{\rm{I}}1}$, $\lambda_{0}$ and $\lambda_{0}'$ as they are most relevant for the low-energy bands.
The resulting effective $4\times 4$ low-energy Hamiltonian, valid for $\mathbf{k}$ vectors close to the $\mathrm{K}$ point, for which
$\gamma_0 f(\mathbf{k}) \ll \gamma_1$ [the structural function $f(\mathbf{k})$ is given by Eq.~(\ref{TB-structural function})], is
\beq
H^{\rm LE}_{\rm eff}(\mathbf{k})=\left(\ba{cccc}
		V-\lambda_{\rm I1}   & \i\lambda_{0}\eta^{+}f(\mathbf{k})  & \gamma_3 f(\mathbf{k}) & \i\bigl(\lambda_{0}\beta^{+}+\lambda_{0}'\beta^{-}\bigr)f^*(\mathbf{k})\\
		-\i\lambda_{0}\eta^{+}f^*(\mathbf{k}) & V+\lambda_{\rm I1}     & 0 	  & \gamma_3 f(\mathbf{k})\\
		\gamma_3 f(\mathbf{k})^* & 0 & -V+\lambda_{\rm I1}  & \i\lambda_{0}'\eta^{-}f(\mathbf{k}) \\
		-\i\bigl(\lambda_{0}\beta^{+}+\lambda_{0}'\beta^{-}\bigr)f(\mathbf{k})  & \gamma_3 f(\mathbf{k})^* & -\i\lambda_{0}'\eta^{-}f^*(\mathbf{k}) & -V-\lambda_{\rm I1}
	\ea\right)
\eeq
with the dimensionless parameters
\beq
\begin{aligned}
\eta^{\pm}&=\frac{(V\mp\Delta)\gamma_0\pm\gamma_1\gamma_4}{V^2+\gamma_1^2-\Delta^2}\,,\\
\beta^{\pm}&=\frac{(V\mp\Delta)\gamma_4\pm\gamma_0\gamma_1}{V^2+\gamma_1^2-\Delta^2}\,.
\end{aligned}
\eeq
\end{widetext}



\end{document}